\documentclass[12pt, a4paper]{amsart}
\usepackage{amsmath, bm}
\usepackage{graphicx,psfrag,epsf}
\usepackage{enumerate}
\usepackage[authoryear]{natbib}
\usepackage{url} % not crucial - just used below for the URL 
\usepackage{amsaddr}
\usepackage[makeroom]{cancel}
\usepackage[normalem]{ulem}

\usepackage{subcaption}
\usepackage{amsfonts, bm, bbm}
\usepackage{amssymb}
\usepackage{amsthm}
\usepackage[makeroom]{cancel}
\usepackage{hhline}
\usepackage{caption}
\usepackage[labelformat=simple]{subcaption}

\usepackage{enumitem}

\usepackage{algorithm}
\usepackage{algpseudocode}
\usepackage{multirow}
\usepackage{arydshln}

\newdimen\figrasterwd
\figrasterwd\textwidth

\usepackage{color}
\usepackage{booktabs}

\newtheorem*{LP}{The Likelihood Principle}

\newtheorem{proposition}{Proposition}[section]
\theoremstyle{definition}

\theoremstyle{remark}

\makeatletter
\newcommand{\pushright}[1]{\ifmeasuring@#1\else\omit\hfill$\displaystyle#1$\fi\ignorespaces}
\newcommand{\pushleft}[1]{\ifmeasuring@#1\else\omit$\displaystyle#1$\hfill\fi\ignorespaces}
\makeatother

\definecolor{brown}{rgb}{0.8, 0.33, 0.1}

\def\by {\bm y}

\def\bc {\bm c}

\def\bbSigma {\mathbf \Sigma}

\def\data {\text{data}}

\def\bz {\bm z}

\def\E {\text{E}}
\def\Var {\text{Var}}
\def\Cov {\text{Cov}}

\def\N {\text{N}}

\def\d {\text{d}}
\def\bone {\bm{1}}
\def\PPOS {\text{PPOS}}
\def\PP {\text{PP}}
\def\CP {\text{CP}}

\def\undertilde#1{\mathord{\vtop{\ialign{##\crcr
$\hfil\displaystyle{#1}\hfil$\crcr\noalign{\kern1.5pt\nointerlineskip}
$\hfil\tilde{}\hfil$\crcr\noalign{\kern1.5pt}}}}}

\makeatletter
\def\paragraph{\@startsection{paragraph}{4}%
  \z@{1ex \@plus1ex \@minus.2ex}{-\fontdimen2\font}%
  {\normalfont\itshape}}
\makeatother

\begin{document}

\def\spacingset#1{\renewcommand{\baselinestretch}%
{#1}\small\normalsize} \spacingset{1}

%%%%%%%%%%%%%%%%%%%%%%%%%%%%%%%%%%%%%%%%%%%%%%%%%%%%%%%%%%%%%%%%%%%%%%%%%%%%%%

\title[On Bayesian Sequential Clinical Trial Designs]{On Bayesian Sequential Clinical Trial Designs}

\author[T. Zhou]{Tianjian Zhou$^1$}
\address{Department of Statistics, Colorado State University \\}
\author[Y. Ji]{Yuan Ji$^2$}
\address{Department of Public Health Sciences, University of Chicago}
\email{$^1$tianjian.zhou@colostate.edu, $^2$yji@health.bsd.uchicago.edu}

\thanks{\mbox{} \vspace{1mm} \\ This is an electronic reprint of the original article published by the New England Statistical Society in The New England Journal of Statistics in Data Science (\url{https://doi.org/10.51387/23-NEJSDS24}). This reprint differs from the original in pagination and typographic detail.}


\begin{abstract}
Clinical trials usually involve sequential patient entry.
When designing a clinical trial, it is often desirable to include a provision for interim analyses of accumulating data with the potential for stopping the trial early.
We review Bayesian sequential clinical trial designs based on posterior probabilities, posterior predictive probabilities, and decision-theoretic frameworks.
A pertinent question is whether Bayesian sequential designs need to be adjusted for the planning of interim analyses.
We answer this question from three perspectives: a frequentist-oriented perspective, a calibrated Bayesian perspective, and a subjective Bayesian perspective.
We also provide new insights into the likelihood principle, which is commonly tied to statistical inference and decision making in sequential clinical trials.
Some theoretical results are derived, and numerical studies are conducted to illustrate and assess these designs.
\end{abstract}

\spacingset{1.45}

\maketitle

\section{Introduction}
\label{sec:intro}

\subsection{Background}

In most clinical trials, patient enrollment is staggered, and patients' data are collected sequentially.
When designing a clinical trial, it is often desirable to include a provision for \textit{interim analyses} of accumulating data with the potential for modifying the conduct of the study \citep{pocock1977group, armitage1991interim}. 
For example, in a randomized-controlled trial, if an interim analysis demonstrates that the investigational drug is deemed superior than the standard of care, the trial could be stopped early on grounds of ethics and trial efficiency \citep{geller1987interim}.
The BNT162b2 COVID-19 vaccine trial is a recent case in which four interim analyses were planned with the possibility for declaring vaccine efficacy before the planned end of the trial \citep{polack2020safety}.
 %\citep{fda2020eua, polack2020safety}.

It is well known that frequentist sequential designs need to be adjusted for the planning of interim analyses to maintain desirable frequentist properties \citep{jennison1990statistical}.
For Bayesian sequential designs, however, there has been some controversy regarding whether similar adjustments are required \citep{ryan2020we}.  Some advocated the necessity of these adjustments (e.g., \citealp{fda2010, fda2019}), while others claimed the opposite (e.g., \citealp{berry1985interim, berry1987interim, harrell2020continuous}).

In this article, we review different perspectives on Bayesian sequential designs and answer the question of whether Bayesian sequential designs need to be adjusted for interim analyses.
Our review is not meant to be comprehensive with regard to methodological details including the type of trial (e.g., single-arm or randomized-controlled), type of outcome (e.g., binary, continuous, or time-to-event), or distributional assumption.
Instead, we focus on the fundamentals of Bayesian sequential designs.
A single-arm trial example (to be introduced in Section \ref{sec:single_arm_example}) will be used throughout to demonstrate these designs, but we present an extension for randomized-controlled trials in Section \ref{sec:rct_extension}. 
We consider early stopping rules for efficacy, as futility stopping does not increase the type I error rate of a design (it actually reduces the type I error rate). Discussion on futility stopping is deferred to Section \ref{sec:discussion}.

There is a rich literature on sequential designs (e.g., \citealp{jennison1990statistical, whitehead1997design, jennison2000group}), but the majority is centered around frequentist approaches.
There are also comprehensive reviews on Bayesian trial designs in general (e.g., \citealp{spiegelhalter1994bayesian, berry2006bayesian, berry2010bayesian}), but most do not extensively address sequential trials.
Lastly, there are many insightful discussions on Bayesian sequential designs, such as \cite{cornfield1966sequential, berry1985interim, berry1987interim, freedman1989comparison, jennison1990statistical, freedman1994and, emerson2007bayesian, harrell2020continuous, ryan2020we, stallard2020comparison}.
However, a systematic review on the fundamentals of Bayesian sequential designs has been lacking, and we attempt to fill this important gap. 
Furthermore, as mentioned earlier, in existing works, different authors seem to have vastly different opinions on how Bayesian sequential designs should be formulated.
It turns out that different authors mean quite different things by ``Bayesian sequential designs need/do not need to be adjusted for interim analyses''.
We aim to disentangle the practical and philosophical implications behind these different perspectives.

Our contributions include the following.  
(i) In Bayesian sequential designs, a pertinent question is whether adjustments for the planning of interim analyses are necessary.
We attempt to answer this question from multiple perspectives. 
From a frequentist-oriented perspective, such adjustments are necessary for achieving desirable frequentist properties such as controlling the type I error rates;
from a calibrated Bayesian perspective, such adjustments may be needed to achieve desirable operating characteristics under plausible scenarios (we will discuss the differences between achieving desirable operating characteristics versus achieving desirable frequentist properties);
lastly, from a subjective Bayesian perspective, such adjustments are unnecessary, and the design only needs to reflect subjective beliefs.
We comment on the three perspectives and make our recommendation.
(ii) We put forward a proposal for a calibrated Bayesian approach to sequential designs.
Specifically, we propose false discovery rate (FDR) and false positive rate (FPR) as potential metrics to evaluate sequential designs.
We derive theoretical results regarding the FDR and FPR of a Bayesian sequential design and present simulation studies to demonstrate the practical usage of the calibrated Bayesian approach.
(iii) We summarize Bayesian sequential designs based on posterior probabilities, posterior predictive probabilities,  and decision-theoretic frameworks.
We discuss the connections between designs using posterior credible intervals and those using formal Bayesian hypothesis testing.
%We also compare between Bayesian and frequentist designs. 
(iv) It is often believed that according to the likelihood principle (LP), decision making in a sequential trial should not depend on unrealized events.
However, our investigation shows that the LP gives little guidance in assessing the overall performance of a decision procedure. 
In particular, the LP does not preclude one from utilizing additional information (including unrealized events) for decision making.
Therefore, our view is that the LP should not be used as an argument for or against Bayesian or frequentist sequential designs.
To illustrate our findings, we present an example of a Bayesian decision-theoretic design in which different decisions will be made based on the same observed data but different interim analysis plans.

\subsection{An Illustrative Example}
\label{sec:single_arm_example}

To illustrate the discussion, consider a single-arm trial that aims to establish the therapeutic effect of an investigational drug. 
Suppose that a total of $K$ analyses, including $(K-1)$ interim analyses and a final analysis, are planned during the course of the trial. 
At the $j$th analysis, data of $n_j$ patients are accumulated, denoted by $y_1, y_2, \ldots, y_{n_j}$ and assumed independently and normally distributed with mean $\theta$ and variance $\sigma^2$.
Here,  $\theta$ is parameterized such that a positive value of $\theta$ is indicative of a therapeutic effect, and $\sigma^2$ is assumed known for simplicity.  
The planned maximum sample size is denoted by $n_K$ and can be determined based on a power requirement or the amount of available resources.
As a simple example, assume patients are enrolled in groups of equal size $g$, thus $n_j = jg$.
If $g = 1$, it leads to the fully sequential case, known as \textit{continuous monitoring}; if $g > 1$, it is called the \textit{group sequential} case, which is more feasible in practice.
The primary research question of the trial can be formulated as the following hypothesis test,
\begin{align}
H_0: \theta \leq 0 \quad \text{vs} \quad H_1: \theta > 0.
\label{eq:test}
\end{align}
At each analysis, the hypothesis test is performed. If certain stopping rule is triggered, say the $z$-statistic $z_j > c_j$ for some stopping boundary $c_j$, $H_0$ is rejected, and the trial is terminated for efficacy. 
Here, 
\begin{align*}
z_{j} = \bar{y}_{j} \cdot \frac{\sqrt{n_j}}{\sigma},  \quad \text{and} \quad \bar{y}_j = \frac{1}{n_j} \sum_{i=1}^{n_j} y_i.  
\end{align*}
This is referred to as \textit{data-dependent} or \textit{optional} stopping. 
When $\sigma$ is unknown, one would replace the $z$-statistics with the corresponding $t$-statistics; little would change in the overall setup.
A question central to sequential designs is the specification of those stopping boundaries. 

\subsection{Overview of Frequentist and Bayesian Sequential Designs}

Frequentist sequential designs are concerned with controlling the overall type I error rate of the sequential testing procedure. 
The type I error rate refers to the probability of falsely rejecting $H_0$ at any analysis (in hypothetical repetitions of the trial), given that $H_0$ is true.
In the single-arm trial example, the maximum type I error rate is attained when $\theta = 0$ and is given by 
\begin{align}
\alpha = \Pr(z_1 > c_1 \text{ or } z_2 > c_2 \text{ or } \cdots \text{ or } z_K > c_K \mid \theta = 0).
\label{eq:type_I_error}
\end{align}
If each test is performed at a constant nominal level, $\alpha$ will inflate as $K$ grows and will eventually converge to 1 as $K \rightarrow \infty$ \citep{armitage1969repeated}.
Therefore, adjustments to the stopping boundaries are necessary to ensure that the type I error rate is maintained at a desirable level.
Examples of such adjustments include the Pocock or O'Brien-Fleming procedure \citep{pocock1977group, o1979multiple}, the error spending approach \citep{slud1982two, gordon1983discrete}, and the stochastic curtailment approach \citep{gordon1982stochastically}.
We provide a brief review of some frequentist sequential designs in Appendix \ref{sec:freq}.

Without accounting for the sequential nature of the hypothesis test, Bayesian designs can suffer the same problem of type I error inflation, which can be unsettling for statisticians who care about controlling the type I error rates. 
Therefore, in many Bayesian sequential trial designs, the stopping boundaries are also determined to control the type I error rate at a desirable level \citep{zhu2017bayesian, shi2019control}.
As an example, the recent BNT162b2 COVID-19 vaccine trial was designed using a Bayesian approach with four planned interim analyses \citep{polack2020safety}. 
The stopping boundaries were chosen such that the overall type I error rate was controlled at 2.5\%. 
Indeed, regulatory agencies generally recommend demonstration of adequate control of the type I error rate for any trial design to be acceptable \citep{fda2010, fda2019}. 
On the other hand, the type I error rate is a frequentist concept, the calculation of which involves an average over unrealized events such as hypothetical repetitions of the trial.
Bayesian inference can be performed based solely on the observed data from the actual (and lone) trial and does not have to be concerned with type I error rate control, since the same trial is not assumed to repeat, hypothetically or in practice.
Some think that the type I error rate is not the quantity that one should pay most attention to \citep{harrell2020pvalues}.
Also, according to the likelihood principle (LP), unrealized events should be irrelevant to the statistical evidence about a parameter \citep{berger1988likelihood}.
Therefore, some Bayesian statisticians have written that the choice of the stopping rules does not need to depend on the planning of interim analyses \citep{berry1985interim, berry1987interim}.
For example, one may stop the trial at any analysis provided that $\Pr(\theta  > 0 \mid \data)$ exceeds some threshold, or if stopping minimizes the posterior expected loss.
We will elaborate on these issues in the upcoming sections.

The remainder of the paper is structured as follows. 
In Section \ref{sec:three_perspectives}, motivated by a sequential design based on posterior probabilities, we summarize the philosophy of Bayesian sequential designs into three categories.
In Section \ref{sec:bayesian_others}, we review selected Bayesian sequential designs based on posterior predictive probabilities and decision-theoretic frameworks. 
In Section \ref{sec:principles}, we comment on the LP, which is commonly tied to statistical inference and decision making in sequential clinical trials.
In Section \ref{sec:example}, we present some numerical studies.
Finally, in Section \ref{sec:discussion}, we conclude and discuss some other considerations including futility stopping rules and two-sided tests.
A brief review of frequentist designs and the proof of the theoretical results are provided in the Appendix.

\section{Three Perspectives on Bayesian Sequential Designs}
\label{sec:three_perspectives}

Consider the single-arm trial in Section \ref{sec:single_arm_example}.
In Bayesian sequential designs, the early stopping rules are typically based on the posterior probability (PP) of $\theta$ being greater than some threshold (e.g., \citealp{thall1994practical, heitjan1997bayesian}).
Assume the time and frequency of interim analyses are given in advance.
Let $\pi(\theta)$ denote the prior distribution of $\theta$.
At analysis $j$, the posterior distribution of $\theta$ is given by Bayes' rule,
\begin{align*}
p(\theta \mid \by_j) = \frac{f(\by_j \mid \theta) \pi(\theta)}{\int f(\by_j \mid \theta) \pi(\theta) \d \theta},
\end{align*}
where $\by_j = (y_1, \ldots, y_{n_j})$ is the vector of accumulating data up to analysis $j$, and $f(\by_j \mid \theta)$ denotes the sampling distribution of $\by_j$.
When the prior for $\theta$ is a conjugate normal distribution, $\theta \sim \N(\mu, \nu^2)$, the above posterior is available in closed form,
\begin{align*}
\theta \mid \by_j \sim \N \left( \frac{\mu \nu^{-2} + \bar{y}_j n_j \sigma^{-2}}{\nu^{-2} + n_j \sigma^{-2}}, \frac{1}{\nu^{-2} + n_j \sigma^{-2}} \right).
\end{align*}
If 
\begin{align}
\PP_j = \Pr(\theta > 0 \mid \by_j) > \gamma_j
\label{eq:bayes_PP_stopping_rule}
\end{align}
for some threshold $\gamma_j$, $H_0$ is rejected, the trial is stopped, and efficacy of the drug is declared.
This is equivalent to 
\begin{align}
z_j > c_j, \quad \text{where} \quad
c_j = q_{1-\gamma_j} \sqrt{1 + \frac{\nu^{-2}}{n_j \sigma^{-2}}} - \frac{\mu \nu^{-2}}{\sqrt{n_j \sigma^{-2}}},
\label{eq:bayes_boundary}
\end{align}
and $q_{1-\gamma_j}$ is the upper $(1-\gamma_j)$ quantile of the standard normal distribution.
It remains to specify the prior $\pi(\theta)$ and threshold values $\{ \gamma_1, \ldots, \gamma_K \}$. 
We present three perspectives next and our comments and recommendation later in Section \ref{sec:comments_perspectives}.

\subsection{The Frequentist-oriented Perspective} 
\label{sec:freq_oriented_perspective}

Without accounting for multiple looks at the data,  the stopping rule in Equation \eqref{eq:bayes_boundary} can lead to type I error rate inflation.
As an example, consider a $\N(0, 1^2)$ prior on $\theta$ and constant threshold values $\gamma_1 = \cdots = \gamma_K = 0.95$.
Suppose the outcome variance $\sigma^2 = 1$, the maximum sample size $n_K = 1000$, and patients are enrolled in equal group sizes.
Using Equation \eqref{eq:type_I_error}, the type I error rates are $\alpha = 0.05, 0.08, 0.13, 0.17, 0.30$, and $0.39$ for $K = 1, 2, 5, 10, 100$, and $1000$, respectively.
Therefore,  due to regulatory guidance \citep{fda2010, fda2019}, one should adjust $\pi(\theta)$ and $\{ \gamma_1, \ldots, \gamma_K \}$ according to the planning of interim analyses to achieve desirable type I error rate control (and possibly other frequentist properties). 
We refer to this as a frequentist-oriented approach.

With an intended type I error rate, the parameters in a Bayesian sequential design can be chosen in multiple ways.
For prespecified threshold values,  type I error rate control can be achieved by using a conservative prior.
\cite{freedman1989comparison} and \cite{freedman1994and} demonstrated that by tuning the prior distribution of $\theta$, one could achieve stopping boundaries similar to or more conservative than Pocock's or O'Brien-Fleming's boundaries.
In our case, we can simply set $\mu = 0$ and adjust $\nu^2$ according to the planning of interim analyses. From Equation \eqref{eq:bayes_boundary}, when $\mu = 0$, the stopping boundaries monotonically increase as $\nu^2$ decreases. 
For example, consider the single-arm trial with an outcome variance of $\sigma^2 = 1$, a maximum sample size of $1000$, $K = 5$ analyses, and equal group sizes.
Then, with threshold values $\gamma_j \equiv 0.95$, a $\N(0, 0.054^2)$ prior for $\theta$ controls the type I error rate at 0.05.
The corresponding stopping boundaries for $z_j$'s are shown in Table \ref{tbl:boundaries}.

Alternatively, for a given prior $\pi(\theta)$,  type I error rate control can be attained by adjusting the threshold values $\{ \gamma_1, \ldots, \gamma_K\}$.
For the single-arm trial example, one may equate the stopping boundaries in Equation \eqref{eq:bayes_boundary} to the corresponding boundaries in any frequentist sequential design.
For example, suppose $\{ c_1, \ldots, c_K \}$ are 
O'Brien-Fleming boundaries, then $\gamma_j$ may be set at
\begin{align*}
\gamma_j = \Phi \left( \frac{c_j + \mu \nu^{-2} / \sqrt{n_j \sigma^{-2}}}{\sqrt{1 + \nu^{-2} / \left( n_j \sigma^{-2} \right)}} \right).
\end{align*}

For more complicated trials (e.g., randomized-controlled, binary outcome),  tuning $\pi(\theta)$ and $\{ \gamma_1, \ldots, \gamma_K\}$ to achieve desirable type I error rate control is more challenging and may require numerical methods. 
See, for example,  \cite{zhu2017bayesian, shi2019control, stallard2020comparison}.

\subsection{The Subjective Bayesian Perspective} 
\label{sec:subjective_bayesian_perspective}

From a subjective Bayesian point of view (see, e.g., \citealp{goldstein2006subjective, robinson2019properties}), the prior $\pi(\theta)$ should be specified to reflect a subjective belief on $\theta$ before the trial, and the threshold values $\{ \gamma_1, \ldots, \gamma_K \}$ should be chosen to represent personal tolerance of risk.
For example, a positive (or negative) prior mean for $\theta$ represents that the investigator's prior belief on the treatment effect is optimistic (or pessimistic). Similarly, the prior variance for $\theta$ reflects the investigator's uncertainty about the prior opinion.
In practice, $\pi(\theta)$ could be elicited from preclinical data and historical clinical trials with a similar setting.
On the other hand, the choice of the threshold values can be justified from a decision-theoretic perspective.
See, e.g., \cite{robert2007bayesian} (Chapter 5.2).
At analysis $j$, the possible decision is denoted by $\varphi_j$, where $\varphi_j = 1$ (or 0) indicates rejecting $H_0$ and stopping the trial (or failing to reject $H_0$ and continuing enrollment if $j < K$).
Assume the loss associated with decision $\varphi_j$ is
\begin{align}
\ell_j(\varphi_j, \theta) = 
\begin{cases}
\xi_{1 j} \cdot \bone(\theta \leq 0), \quad &\text{if $\varphi_j = 1$;} \\
\xi_{0 j} \cdot \bone(\theta > 0), \quad &\text{if $\varphi_j = 0$.}
\end{cases}
\label{eq:loss_naive}
\end{align}
Then, the posterior expected loss of $\varphi_j$ is $L_j(\varphi_j,  \by_j) = \int \ell_j(\varphi_j, \theta) p(\theta \mid \by_j) \d \theta$,
and the decision that minimizes $L_j(\varphi_j,  \by_j)$ is
\begin{align*}
\tilde{\varphi}_j(\by_j) = 
\begin{cases}
1, \quad &\text{if $\Pr(\theta > 0 \mid \by_j) > \frac{\xi_{1 j}}{\xi_{0 j} + \xi_{1 j}}$;} \\
0, \quad &\text{otherwise.}
\end{cases}
\end{align*}
By setting $\gamma_j$ at $\xi_{1 j} / (\xi_{0 j} + \xi_{1 j})$, the stopping rule in Equation \eqref{eq:bayes_boundary} minimizes the posterior expected loss. 
In practice, one could specify the loss function $\ell_j(\varphi_j, \theta)$ based on personal tolerance of risk and then derive the $\gamma_j$'s subsequently.
For example, if one wants to be conservative about rejections early in the trial, one could consider increasing the loss of false rejections at early interim analyses \citep{rosner1995bayesian}.
Of course, the particular loss function in Equation \eqref{eq:loss_naive} is a naive choice and ignores the cost of patient enrollment.
A more stringent way of formulating the loss function should take into account the sequential nature of the trial.
For example, a decision to continue the trial should be made based on balancing the cost of enrolling more patients and the gain of acquiring more information.
More discussion on this point is deferred to Section \ref{sec:decision-theoretic}.

We see that by taking this particular subjective Bayesian approach, one does not need to take frequentist properties into account.
For example, suppose that $\xi_{1 j} = 19 \cdot \xi_{0 j}$ for all $j$, then one can reject $H_0$ and stop the trial at any analysis as long as $\Pr(\theta > 0 \mid \by_j) > 0.95$.
As \cite{edwards1963bayesian} stated, ``it is entirely appropriate to collect data until a point has been proven or disproven, or until the data collector runs out of time, money, or patience.''
This point has also been made by \cite{harrell2020continuous}.

Such a procedure is vulnerable to type I error rate inflation, which would bother many practitioners.
However, it has been argued that the type I error rate is not the quantity that one should pay most attention to \citep{harrell2020pvalues}, because its calculation is conditioned on an assumption rather than something knowable.
Subjective Bayesians argue that what matters is the probability of ``regulator's regret'', $\Pr(\theta \leq 0 \mid \data)$, conditioned on the available data.
Also, the calculation of the type I error rate involves an average over unrealized event that may arise for hypothetical values of $\theta$. However, based on the LP, unobserved events are irrelevant to the evidence about $\theta$ \citep{berry1985interim, berry1987interim}.
We provide more discussion in Section \ref{sec:principles}.

A similar critique on the subjective Bayesian approach is the issue of ``sampling to a foregone conclusion'' \citep{cornfield1966bayesian}.
However, \cite{berry1985interim, berry1987interim} argued that this is not a threat, because the sequence of posterior probabilities, $\{ \Pr(\theta > 0 \mid y_1, \ldots, y_n): n = 1, 2, \ldots \}$, is a martingale.
%Note that $\E [ \Pr(\theta > 0 \mid y_1, \ldots, y_n, y_{n+1}) \mid  y_1, \ldots, y_n ] = \Pr(\theta > 0 \mid y_1, \ldots, y_n)$, where the expectation is taken with respect to the posterior predictive distribution of $y_{n+1}$. Therefore,
%\begin{multline*}
%\E \big[ \Pr(\theta > 0 \mid y_1, \ldots, y_n, y_{n+1}) \mid \Pr(\theta > 0 \mid y_1, \ldots, y_n) \big]   = \\
%\E \Big\{ \E \big[ \Pr(\theta > 0 \mid y_1, \ldots, y_n, y_{n+1}) \mid y_1, \ldots, y_n, \Pr(\theta > 0 \mid y_1, \ldots, y_n) \big] \Big \}\\
% = \Pr(\theta > 0 \mid y_1, \ldots, y_n).
%\end{multline*}
%See Appendix \ref{app:sec:posterior_martingale} for a proof.
If the posterior probability of $\{ \theta > 0 \}$ is less than 0.95 given $n$ observations, say 0.94, then after the next observation, it may increase or decrease with an expected value of 0.94. 
In other words, one cannot guarantee reaching $\Pr(\theta > 0 \mid \data) > 0.95$ with more data.
Specifically, when the sampling distribution of $y_i$'s is normal, the expected number of additional observations required to raise $\Pr(\theta > 0 \mid \data)$ any prescribed amount is infinite.
This is analogous to the expected hitting time of a Brownian motion, which is infinite (see, e.g., Chapter 8.2 in \citealp{ross1996stochastic}).

\subsection{The Calibrated Bayesian Perspective} 
\label{sec:calibrated_bayesian_perspective}

Although Bayesian probabilities represent degrees of belief in some formal sense, for practitioners and regulatory agencies, it can be pertinent to examine the operating characteristics of Bayesian designs in repeated practices. One could calibrate the prior and threshold values in a Bayesian sequential design to achieve desirable operating characteristics under a range of plausible scenarios, and we refer to this as a calibrated Bayesian approach \citep{rubin1984bayesianly, little2006calibrated}.
We provide more background on the calibrated Bayesian perspective in Appendix \ref{sec:app:background_calibrated_Bayes}.

We distinguish between \textit{operating characteristics} and \textit{frequentist properties}:
we use the former to refer to the long-run average behaviors of a statistical procedure in a series of (possibly different) trials, and use the latter to refer to those in (imaginary) repetitions of the same trial.
In other words, operating characteristics represent averages over a joint data-parameter distribution,  while frequentist properties represent averages over a data distribution given a fixed parameter.
See, e.g.,  \cite{rubin1984bayesianly, bayarri2004interplay}.
Frequentist properties are a special class of operating characteristics.

What kinds of operating characteristics could be examined?
Consider the single-arm trial example.
Imagine an infinite series of such trials with true but unknown treatment effects $\{ \theta^{(1)}, \theta^{(2)}, \ldots \}$, which constitute some population distribution $\pi_0(\theta)$.
For each trial, patient outcomes $\by_K \sim f_0(\by_K \mid \theta)$ and are observed sequentially,  where $\by_K = (y_1, \ldots, y_{n_K})$.
Suppose a Bayesian design with stopping rules given by Equation \eqref{eq:bayes_PP_stopping_rule} is applied to every trial with a prior model $\pi(\theta)$, a sampling model $f(\by_K \mid \theta)$, and threshold values $\{ \gamma_1, \ldots,  \gamma_K \}$.
Similar to the rationale of type I error rate control, we propose to control the FDR and FPR of the design  in the infinite series of trials for a range of plausible $f_0(\by_K \mid \theta) \pi_0(\theta)$. 
This is because false rejections of the null may result in continuation of a drug development program that will ultimately fail, increasing the cost associated with the failure.
The FDR is the relative frequency of false rejections among all trials in which $H_0$ is rejected,  and the FPR is the relative frequency of false rejections among all trials with nonpositive treatment effects $\theta$'s. 
Mathematically, let 
\begin{align}
\Gamma = \big \{ \by_K: \exists j \in \{1, \ldots, K\} \; \text{s.t.} 
\Pr(\theta > 0 \mid \by_j) > \gamma_j \; \text{at analysis $j$} \big \}
\label{eq:rej_region}
\end{align}
denote the rejection region of the design.  That is, $H_0$ is rejected if $\by_K \in \Gamma$. Then,
\begin{align*}
\text{FDR}(\pi_0, f_0, \Gamma) &= \frac{\int_{\by_K \in \Gamma} \int_{\theta \leq 0} f_0(\by_K \mid \theta) \pi_0(\theta) \d \theta \d \by_K}{\int_{\by_K \in \Gamma} f_0(\by_K) \d \by_K}, \quad \text{and} \\
\text{FPR}(\pi_0, f_0, \Gamma) &= \frac{\int_{\by_K \in \Gamma}  \int_{\theta \leq 0}  f_0(\by_K \mid \theta) \pi_0(\theta) \d \theta \d \by_K }{\int_{\theta \leq 0} \pi_0(\theta) \d \theta}.
\end{align*}
Our definitions of the FDR and FPR are slightly different from, but closely related to, their typical definitions in a frequentist sense (see, e.g., \citealp{storey2003positive}).

The calibration of the design parameters is typically done through computer simulations.  
For each plausible $f_0(\by_K \mid \theta) \pi_0(\theta)$, one could generate $S$ hypothetical trials with treatment effects $\{ \theta^{(1)}, \theta^{(2)}, \ldots,  \theta^{(S)}\}$ and outcomes $\{ \by_K^{(1)}, \by_K^{(2)}, \ldots,  \by_K^{(S)}\}$ (for some large $S$). 
Then, the FDR and FPR are respectively approximated by
\begin{align}
\begin{split}
\widehat{\text{FDR}} &= \frac{\sum_{s = 1}^S \bm 1 \left ( \by_K^{(s)} \in \Gamma, \theta^{(s)} \leq 0 \right) }{\sum_{s = 1}^S \bm 1 \left ( \by_K^{(s)} \in \Gamma \right)},
\quad \text{and} \\
\widehat{\text{FPR}} &= \frac{\sum_{s = 1}^S \bm 1 \left ( \by_K^{(s)} \in \Gamma, \theta^{(s)} \leq 0 \right) }{\sum_{s = 1}^S \bm 1 \left( \theta^{(s)} \leq 0 \right )}.
\end{split}
\label{eq:fdr_fpr}
\end{align}
The prior and threshold values in the Bayesian design can be chosen such that $\widehat{\text{FDR}}$ and $\widehat{\text{FPR}}$ do not exceed some prespecified levels for every plausible $f_0(\by_K \mid \theta) \pi_0(\theta)$.
Note that the simulations here are different from those for frequentist-oriented approaches. For the latter, hypothetical repetitions of the same trial are simulated with an assumed true treatment effect.

In certain contexts, there are theoretical guarantees on the operating characteristics of Bayesian sequential designs.
Specifically,  the following proposition provides such an example.
\begin{proposition}
\label{prop:fdr_fpr_bound}
Let $\Gamma$ in Equation \eqref{eq:rej_region} represent the rejection region of a Bayesian design.
Assume the joint model for $(\by_K, \theta)$ in the Bayesian design is the same as the actual joint distribution of $(\by_K, \theta)$ in a series of trials, i.e., $f(\by_K \mid \theta) \pi(\theta) = f_0(\by_K \mid \theta) \pi_0(\theta)$.
Then, the FDR and FPR of the Bayesian design are upper bounded regardless of the time ($n_j$'s) and frequency ($K$) of interim analyses,
\begin{align*}
&\text{FDR}(\pi_0, f_0, \Gamma) \leq 1 - \gamma_{\min}, \quad \text{and} \\
&\text{FPR}(\pi_0, f_0, \Gamma) \leq \frac{(1 - \gamma_{\min}) \cdot \int_{\theta > 0} \pi(\theta) \d \theta}{\gamma_{\min} \cdot \int_{\theta \leq 0} \pi(\theta) \d \theta},
\end{align*}
where $\gamma_{\min} = \min \{ \gamma_1, \ldots, \gamma_K \}$.
\end{proposition}
The proof is given in Appendices \ref{sec:app:fdr} and \ref{sec:app:fpr}.
Therefore, from a calibrated Bayesian perspective, the prior on $\theta$ could be elicited to resemble the actual distribution of $\theta$ in repeated practices,
%as closely as possible, 
and the threshold values reflect acceptable FDR and FPR levels.

In general, requiring a design to have good operating characteristics (under plausible scenarios) is more lenient than requiring it to have good frequentist properties (for all possible parameter values).
For example, the type I error rate is essentially the FPR when $\pi_0(\theta)$ is a point mass. Stringent type I error rate requires that the FPR is controlled for \textit{all possible} $\pi_0(\theta)$, even when $\pi_0(\theta)$ is a point mass at 0, while the calibrated Bayesian approach only requires the FPR to be controlled for \textit{plausible} $\pi_0(\theta)$.
In this sense, the calibrated Bayesian approach can be thought of as a middle ground between the frequentist-oriented approach and the subjective Bayesian approach.

\subsection{Our Comments on the Three Perspectives}
\label{sec:comments_perspectives}

We have reviewed three perspectives on Bayesian sequential designs, which are summarized in Table \ref{tbl:three_perspectives}.
 Although the three perspectives seem contradictory,  they are not mutually exclusive. 
For example, if the investigator is conservative about a new drug and is cautious about false rejections, then he/she may take a subjective Bayesian approach with a large loss for a false positive decision.
This can lead to low FDR and FPR, or even a low type I error rate.
In other words, subjective Bayesians may produce desirable operating characteristics for calibrated Bayesians, or desirable frequentist properties for frequentist-oriented Bayesians.

\begin{table}[h!]
\caption{Summary of the three perspectives on Bayesian sequential designs. } 
\label{tbl:three_perspectives}
\begin{center}
\scalebox{0.82}{
\begin{tabular}{lp{3in}p{2.2in}}
\toprule
Perspective & Description & Suitable contexts \\
\midrule
Frequentist-oriented & Specifying design parameters to achieve desirable frequentist properties (e.g.,  type I error rate) \vspace{1mm} & Large-scale confirmatory trials \\
Subjective Bayesian & Specifying design parameters to reflect subjective beliefs and personal tolerance of risk \vspace{1mm} & Trials for rare diseases; pediatric trials for small populations \\
Calibrated Bayesian & Specifying design parameters to achieve desirable operating characteristics (e.g., FDR and FPR) under plausible scenarios & Animal studies for drug screening; early-phase trials (e.g., dose finding) \\
\bottomrule
\end{tabular}
}
\end{center}
\end{table}

In some contexts, a specific approach can be more applicable and acceptable compared to the others.
For example, for large-scale confirmatory trials (e.g., COVID-19 vaccine trials), type I error rate control is enforced by regulators, and thus only the frequentist-oriented perspective is accepted.
Indeed, there are some challenges with the subjective and calibrated Bayesian approaches in those settings. See, e.g., \citep{berry2010bayesian, spiegelhalter1994bayesian}.
With a large number of enrolled patients, a large population that could potentially benefit from the treatment, and multiple decision makers with distinctive prior opinions and tolerances for risk, the process of eliciting costs and benefits can be difficult for subjective Bayesians.
As \cite{spiegelhalter1994bayesian} noted,  ``when the decision is whether or not to discontinue the trial, coupled with whether or not to recommend one treatment in preference to the other, the consequences of any particular course of action are so uncertain that they make the meaningful specification of utilities rather speculative.''
From a calibrated Bayesian perspective, one could elicit the prior for $\theta$ based on historical trials of similar drugs and/or conditions. However, there may be concerns that high or low rates of historical success (e.g., pembrolizumab for solid tumors with a high success rate) may bias the inference for a new trial and trigger incentives for investigators to concentrate clinical research toward attractive areas and selected conditions.
On the other hand, the prior for $\theta$ could also be based on all historical trials regardless of drugs and conditions. However, the distribution of treatment effects can be highly variable over time, and different types of trials have vastly different endpoints, which are difficult to summarize into a common distribution.
As a result, utilization of Bayesian designs for phase III trials
requires a case-by-case discussion that involves extensive examination
of prior elicitation, inference procedures, and simulation results,
which has been highlighted by several guidances from the U.S. Food and Drug Administration \cite{fda2010,fda2019,fda2020}.

The subjective Bayesian perspective can be useful in trials for rare
diseases and pediatric trials for small populations. In those
situations, simple loss functions may be elicited, and prior
distributions can be derived by eliciting expert opinion
\citep{kidwell2022application}. The
elicitation process usually involves interviewing multiple subject
experts such as physicians and their team members, and summaries of
the interviews can be reported in the form of statistics like medians,
modes, and percentiles. Lastly, a prior distribution can be estimated
by fitting a parametric distribution to match the summary
statistics.

Lastly, the calibrated Bayesian perspective is suitable in exploratory settings, such as animal studies for drug screening and early-phase trials (e.g., dose finding).
For those trials, stringent type I error rate control is optional and often at the discretion of the sponsors.
Eliciting the prior for $\theta$ from previous studies and focusing on FDR/FPR control allow an efficient selection of promising drugs for further development.

Influenced by \cite{rubin1984bayesianly, little2006calibrated, robinson2019properties},
our recommendation is to regard the subjective Bayesian paradigm as ideal in principle but often rely on frequentist-type metrics to better communicate Bayesian designs and understand the practical implications of different priors, loss functions, and threshold values.
The LP is sometimes viewed as an argument against the consideration of frequentist-type metrics in hypothetical trials. However, we will demonstrate in Section \ref{sec:principles} that the LP does not preclude one from utilizing frequentist-type metrics to assess a decision procedure.
Still, we advocate the use of operating characteristics under plausible scenarios, in addition to standard frequentist properties, for evaluating trial designs in either exploratory or confirmatory settings.  Metrics like the FDR and FPR have not been used for drug approval, but arguably, they reflect the reality better than frequentist properties. In real life, different clinical trials would have different treatment effects.

\subsection{Bayesian Hypothesis Testing}
\label{sec:Bayesian_hypothesis_testing}

Before moving on to other topics,
we discuss some additional considerations in Bayesian sequential designs.
First, we present a special class of Bayesian designs based on the posterior probability of the alternative hypothesis through formal Bayesian hypothesis testing.
See, e.g., \cite{johnson2009bayesian}.
For the single-arm trial example, to test Equation \eqref{eq:test}, we need to specify the priors for $\theta$ under both the null and alternative hypotheses,
\begin{align*}
\theta \mid H_0 \sim \pi^{(0)}(\theta), \qquad 
\theta \mid H_1 \sim \pi^{(1)}(\theta).
\end{align*}
Importantly,  $\pi^{(0)}(\theta)$ and $\pi^{(1)}(\theta)$ have supports on $(-\infty, 0]$ and $(0, \infty)$, respectively.
Then, the prior probability for each hypothesis is also specified, $\Pr(H_0) = 1 - \omega$ and $\Pr(H_1) = \omega$. 
At analysis $j$, the posterior probability of $H_1$ is 
\begin{multline}
\Pr(H_1 \mid \by_j) = \frac{ \Pr(H_1) f(\by_j \mid H_1)}{\Pr(H_1) f(\by_j \mid H_1) + \Pr(H_0) f(\by_j \mid H_0)} = \\
\frac{\omega \int_{\theta > 0} f(\by_j \mid \theta) \pi^{(1)}(\theta) \d\theta}{\omega \int_{\theta > 0} f(\by_j \mid \theta) \pi^{(1)}(\theta) \d\theta + (1 - \omega) \int_{\theta \leq 0} f(\by_j \mid \theta) \pi^{(0)}(\theta) \d\theta},
\label{eq:BHT}
\end{multline}
which can be used to decide whether to stop the trial early.
For example, if $\Pr(H_1 \mid \by_j) > \gamma_j$, $H_0$ is rejected, and the trial is stopped.
This approach is equivalent to specifying a mixture prior distribution for $\theta$, 
\begin{align*}
\theta \sim \pi(\theta) = (1 - \omega) \cdot \pi^{(0)}(\theta) + \omega \cdot \pi^{(1)}(\theta), 
\end{align*}
and then stop the trial at analysis $j$ if $\Pr(\theta > 0 \mid \by_j) > \gamma_j$. Note that under the mixture prior,
\begin{align*}
\Pr(\theta > 0 \mid \by_j) = \int_{\theta > 0} p(\theta \mid \by_j) \d \theta 
= \frac{\int_{\theta > 0} f(\by_j \mid \theta) \pi(\theta) \d\theta }{\int_{\theta} f(\by_j \mid \theta) \pi(\theta) \d\theta}
 = \eqref{eq:BHT}
\end{align*}
This relationship has been noted by \cite{zhou2021use}.
Although these two approaches are equivalent, when the primary goal is hypothesis testing, the prior for $\theta$ is usually specified as a mixture of two truncated distributions; when the primary goal is parameter estimation, the prior for $\theta$ is usually specified as a single continuous distribution.

A special case is when $H_0$ is a point hypothesis, say when we test $H_0: \theta = 0$ vs $H_1: \theta \neq 0$.
From a hypothesis testing perspective, the prior for $\theta$ should be a mixture of a point mass at $\theta = 0$ (denoted by $\delta_0(\theta)$) and a continuous distribution, $\pi(\theta) = (1-\omega) \delta_0(\theta) + \omega \pi^{(1)}(\theta)$.
Such a prior distribution is rarely used when the primary goal is parameter estimation.
Lastly, \cite{johnson2009bayesian} and \cite{johnson2010use} recommended the use of non-local prior densities, which incorporate a minimally significant separation between the null and alternative hypotheses, for Bayesian hypothesis testing and applications in trial monitoring.

\subsection{Analysis at the Conclusion of a Sequential Trial}
\label{sec:Bayesian_analysis_completion}

From a Bayesian perspective, after a clinical trial has been completed, all the information about $\theta$ is contained in its posterior distribution. 
Let $t$ denote the stopping time of a sequential trial.
For example, based on the stopping rule in Equation \eqref{eq:bayes_boundary}, 
\begin{align*}
t = 
\begin{cases}
\min \{j: z_j > c_j\}, \; &\text{if $\exists j \in \{ 1, \ldots, K\}$ s.t. $z_j > c_j$;} \\
K, \; &\text{if $z_j  \leq c_j$ for all $j$.} \\
\end{cases}
\end{align*}
Then, $\by_t = (y_1, \ldots, y_{n_t})$ is the vector of accumulating data up to the time of stopping. 
At the time of stopping, the posterior distribution of $\theta$ is given by
\begin{align*}
p(\theta \mid \by_t) = \frac{f(\by_t \mid \theta) \pi(\theta)}{\int_{\theta} f(\by_t \mid \theta) \pi(\theta) \d \theta}.
\end{align*}
One may be worried that the stopping time $t$ is not included in the conditional of $p(\theta \mid \by_t)$. However, assuming that $\theta$ and $t$ are independent conditional on $\by_t$, we have
\begin{align*}
p(\theta \mid t, \by_t) = \frac{f(t, \by_t \mid \theta) \pi(\theta)}{\int_{\theta} f(t, \by_t \mid \theta) \pi(\theta) \d \theta} =
\frac{\cancel{f(t \mid \by_t)} f(\by_t \mid \theta) \pi(\theta)}{\int_{\theta} \cancel{f(t \mid \by_t)} f(\by_t \mid \theta) \pi(\theta) \d \theta} = p(\theta \mid \by_t),
\end{align*}
because $f(t, \by_t \mid \theta) = f(t \mid \by_t, \theta) f(\by_t \mid \theta) = f(t \mid \by_t) f(\by_t \mid \theta)$.
Most often (and in all the designs that we have reviewed), $\theta$ affects $t$ only through the observations $\by_t$, in which case the conditional independence assumption is satisfied, the equation holds,  and the stopping rule plays no role in the posterior distribution of $\theta$. 
See, e.g., \cite{hendriksen2021optional}. 
However, we note that in some situations, $\theta$ could affect $t$ other than just via $\by_t$. For example, if an interim analysis happens because an external trial found a positive treatment effect, which is more likely if $\theta$ is positive and large, this would affect $t$ via external data other than via the current data.

The posterior mean, $\E(\theta \mid \by_t)$, is a commonly used point estimator for $\theta$.
On the other hand, a $100(1 - \alpha)\%$ credible interval for $\theta$ can be constructed as $(\theta^\text{L}, \theta^\text{U})$, where $\theta^\text{L}$ and $\theta^\text{U}$ are the lower and upper $(\alpha/2)$ quantiles of $p(\theta \mid \by_t)$, respectively.
This credible interval has its asserted coverage in repeated practices if the model specification is correct (see Appendix \ref{sec:app:background_calibrated_Bayes}), but the coverage may deteriorate in the presence of model misspecification.
Lastly, the posterior probability of the alternative hypothesis, $\Pr(\theta > 0 \mid \by_t)$, is also reported.

\subsection{Randomized-controlled Trial and Minimum Clinically Important Difference}
\label{sec:rct_extension}

So far, we have been using a single-arm trial to illustrate the designs. In practice, multi-arm trials such as randomized-controlled trials are also very common. 
We briefly outline an extension of the designs for a randomized-controlled trial.
For simplicity, assume the trial outcomes are normally distributed.
At analysis $j$, observed data are $y_{r1}, y_{r2}, \ldots, y_{rn_{rj}} \sim \N(\theta_r, \sigma_r^2)$ for arm $r$, where $r = 1$ and $0$ represent the investigational drug and control arms, respectively. The goal may be to test
\begin{align*}
H_0: \theta_1 - \theta_0 \leq 0 \quad \text{vs} \quad H_1: \theta_1 - \theta_0 > 0.
\end{align*}
Assume $\sigma_1^2$ and $\sigma_0^2$ are known. 
One can specify a prior distribution for $\theta = \theta_1 - \theta_0$, say $\theta \sim \N(\mu, \nu^2)$. The posterior distribution of $\theta$ at analysis $j$ is given by
\begin{multline*}
\theta \mid \by_{1j}, \by_{0j} \sim \N \Bigg[ \frac{\mu \nu^{-2} + (\bar{y}_{1j} - \bar{y}_{0j}) (\sigma_1^2 / n_{1j} + \sigma_0^2 / n_{0j})^{-1}}{\nu^{-2} +  (\sigma_1^2 / n_{1j} + \sigma_0^2 / n_{0j})^{-1} }, \\
 \frac{1}{\nu^{-2} +  (\sigma_1^2 / n_{1j} + \sigma_0^2 / n_{0j})^{-1}} \Bigg],
\end{multline*}
where $\bar{y}_{rj} = \frac{1}{n_{rj}} \sum_{i = 1}^{n_{rj}} y_{ri}$.
Then, one can proceed similarly as before.
An alternative approach is to specify independent priors separately for $\theta_1$ and $\theta_0$ and then use these to obtain a posterior distribution of $\theta$. This will lead to slightly different designs. See \cite{stallard2020comparison}.
When $\sigma_1^2$ and $\sigma_0^2$ are unknown, one needs to specify priors for these parameters as well and calculate the marginal posterior distribution of $\theta$.

In some trials, such as proof-of-concept trials, it may be of interest to evaluate the evidence of the treatment effect being greater than a minimum clinically important difference, denoted by $\Delta$ \cite{chuang2011role, fisch2015bayesian}.  In this case, one may 
replace the stopping rule in Equation \eqref{eq:bayes_PP_stopping_rule} by
\begin{align}
\Pr(\theta > \Delta \mid \by_j) > \gamma^{\Delta}_j.
\label{eq:bayes_PP_stopping_rule2}
\end{align}
Alternatively, the efficacy stopping rule can be based on both Equations \eqref{eq:bayes_PP_stopping_rule} and \eqref{eq:bayes_PP_stopping_rule2}. 
Here,  Equation \eqref{eq:bayes_PP_stopping_rule} speaks to ``does the drug work at all'', while Equation \eqref{eq:bayes_PP_stopping_rule2} addresses ``does the drug have a clinically relevant effect''.
In proof-of-concept trials, Equation \eqref{eq:bayes_PP_stopping_rule2} may be a necessary criterion for a drug to be promoted into full development \cite{fisch2015bayesian}.

\subsection{Comparison with Frequentist Sequential Designs}
\label{sec:Bayes_discussion}

Compared to their frequentist counterparts, Bayesian designs involve additional complexities such as prior elicitation and computational challenges when the posterior distribution is not analytically tractable. 
Still, Bayesian designs have certain advantages (see, e.g., \citealp{freedman1994and}).
First, with a chosen probability model, the data affect posterior inference only through the likelihood function.
In this way, Bayesian inference obeys the LP (\citealp{gelman2013bayesian}, p. 7).
This can be philosophically appealing.
Frequentist inference, on the other hand, may be affected by unrealized events.
We will elaborate on this point in Section \ref{sec:principles}.
Second, the stopping rule of an experiment is irrelevant to the construction and interpretation of a Bayesian credible interval.
In contrast,  a frequentist interval estimate of treatment effect following a group sequential trial crucially depends on the stopping rule.
As \cite{freedman1994and} pointed out, such an interval may be quite unintuitive. Depending on the choice of sample space ordering, the interval may not always include the sample mean and can include zero difference even for data that lead to a recommendation to stop the trial at the first interim analysis (see \citealp{rosner1988exact}).
Third, stringent frequentist inference can be challenging or unsatisfactory if the prescribed stopping rule is not followed.
For example, a trial may be stopped due to unforeseeable circumstances such as the outbreak of COVID-19;
in some cases, it may be desirable to extended a trial beyond the planned sample size.
Some have criticized that the relevance of stopping rules makes it almost impossible to conduct any frequentist inference in a strict sense \citep{berger1980statistical, berry1985interim, berger1988likelihood, wagenmakers2007practical}.
Oftentimes, statisticians are presented with a dataset without knowing how the stopping of the study was decided and why the study was not stopped earlier.
Both factors can affect the frequentist properties of a statistical procedure, while in practice it is infeasible to keep track of them. 
Lastly, when reliable historical information is available, it can be formally incorporated into the design and analysis of the current trial via Bayesian methods. This may lead to improvements in trial efficiency in terms of higher power and saving in sample size (see \citealp{shi2019control}).

\section{Other Types of Bayesian Sequential Designs}
\label{sec:bayesian_others}

\subsection{Designs Based on Posterior Predictive Probabilities}
\label{sec:bayesian_ppp}

In the upcoming sections, we review some other types of Bayesian sequential designs whose early stopping rules are not directly based on $\Pr(\theta > 0 \mid \by_j) > \gamma_j$. 
Similar to the idea of stochastic curtailment \citep{gordon1982stochastically}, posterior predictive probabilities can be used to determine whether to stop a trial early.
See, e.g., 
\cite{dmitrienko2006bayesian, lee2008predictive, saville2014utility}.
Suppose that at the final analysis, efficacy of the drug will be declared if $\Pr(\theta > 0 \mid \by_K) > 1 - \eta$.
At analysis $j \in \{ 1, \ldots, K-1 \}$, the posterior predictive distribution of future observations $\by_{j, K}^* = (y_{n_j+1}^*, \ldots, y_{n_K}^*)$ is
\begin{align*}
p(\by_{j, K}^* \mid \by_j) = \int_{\theta} f(\by_{j, K}^* \mid \theta) p(\theta \mid \by_j) \d \theta,
\end{align*}
and the posterior predictive probability of success (PPOS) is
\begin{align*}
\PPOS_j = \int_{\by_{j, K}^*} \bone \left[ \Pr \left(\theta > 0 \mid \by_j, \by_{j, K}^* \right) > 1 - \eta \right] \cdot p(\by_{j, K}^* \mid \by_j) \d \by_{j, K}^*.
\end{align*}
One may stop the trial early if $\PPOS_j > \gamma_j$ for some threshold $\gamma_j$.
To specify the prior for $\theta$ and the threshold values $\{ \gamma_1, \ldots, \gamma_{K-1}\}$ and $\eta$, one may take one of the approaches in Sections \ref{sec:freq_oriented_perspective}--\ref{sec:calibrated_bayesian_perspective}.

For the single-arm trial example, we have 
\begin{align*}
\bar{y}_{j, K}^* \mid \by_j \sim \N \Bigg( \frac{\mu \nu^{-2} + \bar{y}_j n_j \sigma^{-2}}{\nu^{-2} + n_j \sigma^{-2}}, 
\frac{1}{\nu^{-2} + n_j \sigma^{-2}} + \frac{1}{(n_K - n_j) \sigma^{-2}} \Bigg),
\end{align*}
where $\bar{y}_{j, K}^* = \left( y_{n_j+1}^* + \cdots + y_{n_K}^* \right) / (n_K - n_j)$.
The criterion $\Pr \left(\theta > 0 \mid \by_j, \by_{j, K}^* \right) > 1 - \eta$ is equivalent to 
\begin{align*}
\bar{y}_K^* = \frac{1}{n_K} \left[ n_j \bar{y}_j + (n_K - n_j) \bar{y}_{j, K}^* \right] > 
q_{\eta} \cdot \frac{\sqrt{\nu^{-2} + n_K \sigma^{-2}}}{n_K \sigma^{-2}} - \frac{\mu \nu^{-2}}{n_K \sigma^{-2}}.
\end{align*}
Finally, it can be derived that 
\begin{multline*}
\PPOS_j = 1 - \Phi \Bigg\{ \left[ \frac{1}{\nu^{-2} + n_j \sigma^{-2}} + \frac{1}{(n_K - n_j) \sigma^{-2}} \right]^{-1/2} \cdot \\
\Bigg[ \frac{n_K}{n_K - n_j} \cdot \Bigg( q_{\eta} \cdot \frac{\sqrt{\nu^{-2} + n_K \sigma^{-2}}}{n_K \sigma^{-2}} - 
\frac{\mu \nu^{-2}}{n_K \sigma^{-2}} -
 \frac{\bar{y}_j n_j}{n_K}  \Bigg) - 
\frac{\mu \nu^{-2} + \bar{y}_j n_j \sigma^{-2}}{\nu^{-2} + n_j \sigma^{-2}} 
 \Bigg] \Bigg\}.
\end{multline*}
The PPOS depends on $\eta$ and $n_K$. 
In general, the stopping rules based on PPOS and PP are different, although for given $\eta$ and $n_K$,  one may select $\gamma_j'$ such that $\{ \PPOS_j > \gamma_j' \}$ and $\{ \PP_j > \gamma_j \}$ are equivalent.
As a result, one may also impose type I error rate control on PPOS stopping rules based on the arguments in Section \ref{sec:freq_oriented_perspective}.
As noted by \cite{saville2014utility}, if at the $j$th interim analysis, the amount of data remain to be collected ($n_K - n_j$) is infinity, then $\PPOS_j = \PP_j$ regardless of $\eta$.
Typically, the PPOS is close to the PP at the beginning of a trial and moves toward either 0 or 1 as the trial nears completion.

\subsection{Decision-theoretic Designs}
\label{sec:decision-theoretic}

As described in Section \ref{sec:subjective_bayesian_perspective}, the decisions in a sequential clinical trial can be made by minimizing the expected loss under a decision-theoretic framework.
This approach has been considered by \cite{berry1988one, lewis1994group, stallard1999decision, ventz2015bayesian}, among others.
The idea is that, at each interim analysis, the decision to stop the trial early and reject $H_0$ is associated with some loss if the decision is wrong.
On the other hand, continuing the trial results in more cost in terms of patient recruitment. But with more data, the chance of making a wrong decision may be decreased.
By considering both factors, decision-theoretic designs combine the strengths of designs based on posterior and posterior predictive probabilities.

We illustrate the idea of decision-theoretic designs through the single-arm trial example. 
Let $\varphi_j$ denote a possible decision at analysis $j$. For $j = 1, \ldots, K-1$, $\varphi_j = 1$ (or 0) represents rejecting $H_0$ and stopping the trial early (or failing to reject and continuing enrollment).
For $j = K$, $\varphi_K = 1$ (or 0) represents rejecting (or failing to reject) $H_0$ at the final analysis, and the trial is stopped in either case.
Let $\ell_j(\varphi_j,\theta, \by_j)$ denote the loss of making decision $\varphi_j$ at analysis $j$ given parameter $\theta$ and data $\by_j$.
The posterior expected loss is then $L_j(\varphi_j, \by_j) = \int_{\theta} \ell_j(\varphi_j,\theta, \by_j) p(\theta \mid \by_j) \d \theta$. The optimal decision is $\tilde{\varphi}_j(\by_j) = \arg\min_{\varphi_j} L_j(\varphi_j, \by_j)$ and the associated expected loss is $\tilde{L}_{j}(\by_j) = \min_{\varphi_j} L_j(\varphi_j, \by_j)$, i.e., the Bayes risk.

Suppose that the loss of making decision $\varphi_j = 1$ at analysis $j$ ($j = 1, \ldots, K-1$) is 
\begin{align}
\ell_j (\varphi_j = 1,  \theta, \by_j) = 
\xi_{1j} \cdot \bone(\theta \leq 0),
\label{eq:loss_interim1}
\end{align}
where $\xi_{1j}$ is the loss of mistakenly rejecting $H_0$ and stopping the trial if $\theta \leq 0$.
On the other hand, if $\varphi_j = 0$, the trial continues, $(n_{j+1} - n_j)$ patients will be enrolled until the next analysis, and we assume a unit loss for recruiting each patient. We have
\begin{align}
\ell_j (\varphi_j = 0,  \theta, \by_j) =  \left( n_{j+1} - n_j \right) + 
\int_{\by_{j, j+1}^*} \tilde{L}_{j+1}(\by_{j}, \by_{j, j+1}^*) p(\by_{j, j+1}^* \mid \by_{j}) \d \by_{j, j+1}^*.
\label{eq:loss_interim2}
\end{align}
Here, $\int_{\by_{j, j+1}^*} \tilde{L}_{j+1}(\by_{j}, \by_{j, j+1}^*) p(\by_{j, j+1}^* \mid \by_{j}) \d \by_{j, j+1}^*$ is the Bayes risk at analysis $(j+1)$ marginalized over the posterior predictive distribution on $\by_{j, j+1}^* = (y_{n_j+1}^*, \ldots, y_{n_{j+1}}^*)$, that is, the observations between analyses $j$ and $j+1$.

We also assume the loss of making decision $\varphi_K$ at the final analysis is 
\begin{align*}
\ell_K(\varphi_K, \theta, \by_K) = 
\begin{cases}
\xi_{1K} \cdot \bone(\theta \leq 0), \quad &\text{if $\varphi_K = 1$;} \\
\xi_0 \cdot \bone(\theta > 0), \quad &\text{if $\varphi_K = 0$.}
\end{cases}
\end{align*}
Here, $\xi_{1K}$ is the loss of mistakenly rejecting $H_0$ at the final analysis if $\theta \leq 0$ (a type I error), and $\xi_0$ is the loss of failing to reject $H_0$ if $\theta > 0$ (a type II error).

At analysis $j$, the optimal decision $\tilde{\varphi}_j(\by_j)$ can be solved by backward induction (\cite{degroot1970optimal}, Chapter 12).
First, we calculate $\tilde{L}_K(\by_K)$ for all possible data $\by_K$ that can arise at the final analysis.
Next, using Equations \eqref{eq:loss_interim1} and \eqref{eq:loss_interim2}, we can calculate $\tilde{L}_{K-1}(\by_{K-1})$ for all possible data $\by_{K-1}$ that can arise at analysis $(K-1)$.
Proceeding backward in this way gives $\tilde{L}_{K-2}(\by_{K-2}), \ldots, \tilde{L}_{j}(\by_{j})$.
This procedure requires many minimizations and integrations which may not be analytically tractable. Simulation-based approaches have been proposed to mitigate these computational challenges \citep{muller2007simulation}.

\cite{lewis1994group} demonstrated that by tuning the loss functions, decision-theoretic designs can achieve desirable type I error rate control.
\cite{ventz2015bayesian} considered constrained optimal designs with explicit frequentist requisites. 
Alternatively, the loss functions and prior can be chosen by taking the subjective or calibrated Bayesian approach.

We summarize in Table \ref{tbl:designs} the various methods and measures that give rise to different types of sequential designs, including frequentist designs reviewed in Appendix \ref{sec:freq}.

\begin{table}[h!]
\caption{Summary of methods and measures that give rise to different types of sequential designs.} 
\label{tbl:designs}
\begin{center}
\scalebox{0.78}{
\begin{tabular}{lp{2.7in}p{2.3in}}
\toprule
Method/measure & Stopping criteria for efficacy & Design parameters \\
\midrule
\textbf{Bayesian designs}: & &  \vspace{1mm}\\
Posterior probability & Posterior probability (PP) of drug being efficacious exceeds a prespecified threshold &  Prior for treatment effect; PP thresholds at interim and final analyses \vspace{1mm}\\
Posterior predictive probability & Posterior predictive probability of trial success (PPOS) exceeds a prespecified threshold & Prior for treatment effect; PP threshold at final analysis; PPOS thresholds at interim analyses  \vspace{1mm}\\
Decision-theoretic & Efficacy stopping minimizes posterior expected loss for a prespecified loss function &  Prior for treatment effect; loss functions associated with possible decisions \vspace{1mm}\\
\midrule
\textbf{Frequentist designs}: & &  \vspace{1mm}\\
Frequentist group sequential & Test statistic exceeds a prespecified stopping boundary & Stopping boundaries for test statistics that define a critical region \vspace{1mm}\\
Stochastic curtailment & Conditional power (CP) of trial success, given a hypothetical treatment effect, exceeds a prespecified threshold & Critical value for test statistic at final analysis; CP thresholds at interim analyses  \\
\bottomrule
\end{tabular}
}
\end{center}
\end{table}

\section{The Likelihood Principle}
\label{sec:principles}

Statistical inference and decision making in sequential clinical trials are typically tied to the LP.
We provide some discussions in this section.

Let $Y$ denote a random variable with density $f_\theta(y)$. The likelihood function for $\theta$, given the observed outcome $y$ of the random variable $Y$, is $L_y(\theta) = f_\theta(y)$. That is, the density evaluated at $y$ and considered as a function of $\theta$.
The (strong) LP, as in \cite{birnbaum1962foundations} and \cite{berger1988likelihood}, can be summarized as follows:

\begin{LP}
All the statistical evidence about $\theta$ arising from an experiment is contained in the likelihood function for $\theta$ given $y$. Two likelihood
functions for $\theta$ (from the same or different experiments) contain the same statistical evidence about $\theta$ if they are proportional to one another.
\end{LP}

\cite{birnbaum1962foundations} showed that the LP can be deduced from two widely accepted principles: the sufficiency principle and the conditionality principle. There have been debates regarding Birnbaum's proof and the validity of the LP in general.
A detailed treatment of the LP is outside the scope of this paper.
We refer interested readers to \cite{berger1988likelihood, robins2000conditioning, evans2013does, mayo2014birnbaum, gandenberger2015new, pena2017note}.

What would be the consequences if we accept the LP?
Since the LP deals only with the observed $y$, data that did not obtain and experiments not carried out have no impact on the evidence about $\theta$  \citep{berry1987interim, berger1988likelihood}.
Also, as in \cite{berger1988likelihood},  the LP implies that the reason for stopping an experiment (the stopping rule) should be irrelevant to the evidence about $\theta$.
In a clinical trial, the implication is that early stopping would not affect the evidential meaning of the trial outcome.

As an illustration, consider the example given by \cite{berry1987interim}.
Imagine that a single-arm trial as described in Section \ref{sec:single_arm_example} has been conducted, and 200 outcomes have been recorded that result in a $z$-statistic of $z_1 = 1.75$.
These results are being reported by two investigators A and B, who used the same probability model (including the prior model for $\theta$, if they were to take a Bayesian approach) but had different plans about the next step.
Investigator A planned a second stage for the trial to enroll 200 more patients should it happen that $z_1 \leq 1.88$ (the Pocock stopping boundary, see \cite{pocock1977group}), while investigator B did not plan to enroll any more patients.
According to the LP,  the evidence about $\theta$ provided by the 200 observations is not affected by the investigators' plans. 
%This is intuitive, as every aspect of the study and the data are identical.

Although the LP seems compelling, it has been a source of controversy.
Under the Bayesian paradigm, for any specified prior distribution for $\theta$, if the likelihood functions are proportional as functions of $\theta$, the resulting posterior densities for $\theta$ are identical.
In this sense, Bayesian inference conforms to the LP (\citealp{bernardo2000bayesian}, p. 249; \citealp{gelman2013bayesian}, p. 7).
On the other hand, the LP seems to be incompatible with many frequentist procedures.
In the previous example,  investigator A cannot claim statistical significance using the Pocock design after 200 observations (and may fail again after all 400 observations), while investigator B can using a fixed design with 200 patients ($z_1 > q_{0.05} = 1.645$). 
In other words, these investigators can reach completely different conclusions about the effectiveness of the drug with the exact same data.

\begin{figure}[h!]
\begin{center}
\includegraphics[width = .8\textwidth]{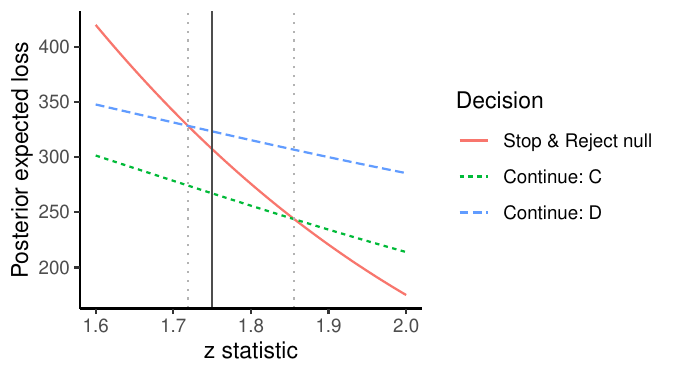}
\end{center}
\caption{Posterior expected losses, as functions of the $z$-statistic, for possible decisions that can be made by investigators C and D at an interim analysis after 200 observations. The trial has a maximum sample size of 400 patients.  Investigator C planned another interim analysis after 300 observations, while investigator D did not plan to conduct any additional interim analysis. The solid vertical line represents an observed $z$-statistic of 1.75 at the interim analysis. The optimal decisions for investigators C and D are continuing enrollment and stopping the trial, respectively.}
\label{fig:LP_example}
\end{figure}

The conflict here does not mean we have to either reject  the LP or reject frequentist procedures.
Explained previously (e.g., \citealp{berger1988likelihood, gandenberger2015two, gandenberger2017differences}), 
the LP is not a decision procedure and gives little guidance in assessing the overall performance of a decision procedure.
The LP implies that only the observed data are relevant to the evidence about $\theta$, but the consequences for making a specific decision may depend on other aspects of an experiment.
First, while the evidence about $\theta$ is trial-specific,  a decision procedure is applied to many trials.
For example, from a regulatory agency's perspective, the action to approve a drug reflects not only the consequences of administering this drug to patients, but also the downstream consequences of that decision rule for other drugs in the future \citep{gandenberger2017differences}.
Therefore, frequentist measures such as the type I error rate can be factored into the decision procedure.
Second, even for a single trial, 
it is not unreasonable to associate the consequences of a decision with unrealized data patterns. 
For example, in a Bayesian sequential design based on posterior predictive probabilities (Section \ref{sec:bayesian_ppp}), the calculation of the PPOS involves an average over the posterior predictive distribution of future data. Such averaging is also required in a Bayesian decision-theoretic design (Section \ref{sec:decision-theoretic}) when calculating the posterior expected loss of a decision based on backward induction.
Imagine an ongoing clinical trial with a maximum sample size of 400 patients and an outcome variance of $\sigma^2 = 1$.
Suppose the Bayesian decision-theoretic design in Section \ref{sec:decision-theoretic} is used.
After 200 outcomes have been recorded, an interim analysis is being performed by two investigators C and D, who used the same probability model with a $\N(0, 1^2)$ prior on $\theta$ but had different plans.
Investigator C planned another interim analysis after 300 observations, while investigator D did not plan to conduct any additional interim analysis.
Suppose the $z$-statistic at the interim analysis is $z_1 = 1.75$.
Then, using the design and loss functions described in Section \ref{sec:decision-theoretic} with $\xi_0 = 400$ and $\xi_{1j} \equiv 19 \xi_0$ for all $j$, the optimal decisions for investigators C and D are continuing enrollment and stopping the trial, respectively.
Specifically, Figure \ref{fig:LP_example} shows the posterior expected losses for possible decisions that can be made by the two investigators. We can see that the existence of a planned future interim analysis has an impact on the posterior expected loss associated with continuing the trial. 
In summary, if a dichotomous decision must be made, the LP does not preclude one from utilizing other information in addition to the observed data.
Therefore, our view is that the LP should not be used as an argument for or against Bayesian or frequentist sequential designs.

Still, the conflict does suggest that if we accept the LP, then frequentist measures such as type I/II error rates and $p$-values may not be used as measures of statistical evidence for or against a hypothesis in a clinical trial \citep{berger1988likelihood}.
This point has been raised by many others as well.
For example, \cite{royall1997statistical} stated that ``Neyman-Pearson statistical theory is aimed at finding good rules for choosing 
from a specified set of possible actions. It does not address the problem of
representing and interpreting statistical evidence, and the decision rules derived from Neyman-Pearson theory are not appropriate tools for interpreting data as evidence.''
In summary, in an ideal world, one may use frequentist measures to design a trial.
However, when reporting statistical analyses results as evidence after trial completion, Bayesian measures that conform the LP should be preferred.

It should also be noted that not all Bayesian procedures are in compliance with the LP.
For example, eliciting the prior for $\theta$ based on the sampling plan, such as using the Jeffreys prior \citep{jeffreys1946invariant}, results in violation of the LP (\citealp{berger1988likelihood}, p. 21).
We have mentioned in Section \ref{sec:freq_oriented_perspective} that one may control the type I error rate of a Bayesian sequential design by calibrating the prior or threshold values. 
To avoid violation of the LP, however, we recommend taking the latter approach and not selecting the prior based on trial planning. 
Intuitively, changing the threshold values only affects  decision making, while changing the prior affects both the evidence about $\theta$ (e.g., point and interval estimations) and decision making.

\section{Numerical Studies}
\label{sec:example}

\subsection{Illustration of the Frequentist-oriented Approach}

As an illustration of the frequentist-oriented approach, we calculate the stopping boundaries for  the $z$-statistics given by some of the aforementioned Bayesian sequential designs with the type I error rate controlled at $\alpha = 0.05$. That is, we compute the $\{c_1, \ldots, c_K \}$ values for which we would stop the trial at analysis $j$ if $z_j > c_j$.
We consider the single-arm trial example described in Section \ref{sec:single_arm_example}. Suppose that a total of $K = 5$ (interim and final) analyses are planned, the maximum sample size is $n_K = 1000$, and patients are enrolled in groups of size 200 ($n_j = 200j$).
The variance for the outcomes is set at $\sigma^2 = 1$ and is assumed known. 
Specifically:
\begin{enumerate}[noitemsep,nolistsep,leftmargin=.25in, label=(\roman*)]
\item For stopping boundaries based on posterior probabilities (Equation \ref{eq:bayes_boundary}), we consider the following two versions. In the first version, we use $\gamma_j \equiv 0.95$ and find that a $\N(0, 0.054^2)$ prior for $\theta$ leads to $\alpha = 0.05$. In the second version, we place a $\N(0, 1^2)$ prior on $\theta$ and find that setting $\gamma_j \equiv 0.983$ leads to $\alpha = 0.05$.
\item For stopping boundaries based on posterior predictive probabilities (Section \ref{sec:bayesian_ppp}), we set $\gamma_j \equiv 0.8$, $\eta = 0.05$, and find that a $\N(0, 0.063^2)$ prior for $\theta$ leads to $\alpha = 0.05$. 
\item For the Bayesian decision-theoretic design (Section \ref{sec:decision-theoretic}), we place a $\N(0, 1^2)$ prior on $\theta$, use $\xi_0 = 1000$, and find that setting $\xi_{1j} \equiv 34890$ leads to $\alpha = 0.05$.
\end{enumerate}

The stopping boundaries are summarized in Table \ref{tbl:boundaries}. 
For comparison, we also include the stopping boundaries produced by the Pocock and O'Brien-Fleming procedures \citep{pocock1977group, o1979multiple} and the linear error spending function \citep{kim1987design}.
See Appendices \ref{sec:pocock} and \ref{sec:error_spending} for more details. 
With $\gamma_j \equiv 0.95$ and a conservative prior $\N(0, 0.054^2)$, the Bayesian design based on posterior probabilities leads to stopping boundaries that lie between Pocock's and O'Brien-Fleming's boundaries; with a $\N(0, 1^2)$ prior and $\gamma_j \equiv 0.983$, it gives stopping boundaries that are similar to Pocock's boundaries.
The Bayesian design based on predictive probabilities with a conservative prior $\N(0, 0.063^2)$ gives boundaries that lie between Pocock's and O'Brien-Fleming's boundaries.
Lastly, by tuning the loss functions, the Bayesian decision-theoretic design leads to stopping boundaries similar to those given by the linear error spending function.

\begin{table}[h!]
\caption{Stopping boundaries for the $z$-statistics given by several Bayesian and frequentist sequential designs. The single-arm trial in Section \ref{sec:single_arm_example} is considered with $K = 5$ analyses, a maximum sample size of $n_K = 1000$, and equal group sizes ($n_j = 200j$).  The design parameters are calibrated such that the type I error rate at $\theta = 0$ is $\alpha = 0.05$ for every design.} 
\label{tbl:boundaries}
\begin{center}
\begin{tabular}{lccccc}
\toprule
Analysis & 1 & 2 & 3 & 4 & 5  \\
No. of patients & 200 & 400 & 600 & 800 & 1000 \\
\midrule
\textbf{Bayesian designs:} \vspace{1mm} \\
Post. prob. (ver. 1) & 2.71 & 2.24 & 2.06 & 1.97 & 1.91 \\
Post. prob. (ver. 2) & 2.13 & 2.12 & 2.12 & 2.12 & 2.12 \\
Post. pred. prob. & 2.50 & 2.26 & 2.18 & 2.11 & 1.84 \\
Decision-theoretic & 2.33 & 2.22 & 2.15 & 2.09 & 1.91 \\
\midrule
\textbf{Frequentist designs:} \vspace{1mm} \\
Pocock & 2.12 & 2.12 & 2.12 & 2.12 & 2.12 \\
O'Brien-Fleming & 3.92 & 2.77 & 2.26 & 1.96 & 1.75 \\
Linear error spending &  2.33 & 2.22 & 2.12 & 2.03 & 1.96 \\
\bottomrule
\end{tabular}
\end{center}
\end{table}

\begin{figure}[ht!]
\begin{center}
\includegraphics[width = .83\textwidth]{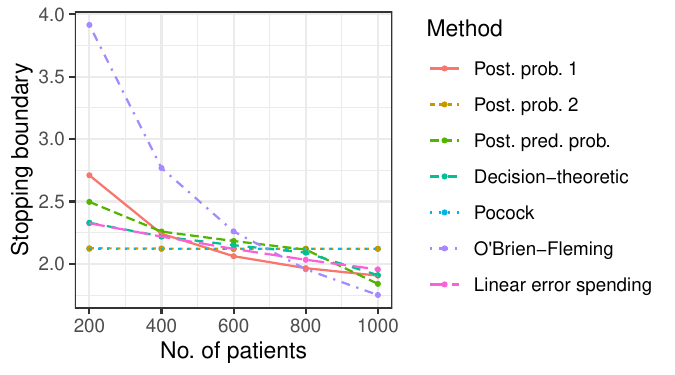}
\includegraphics[width = .83\textwidth]{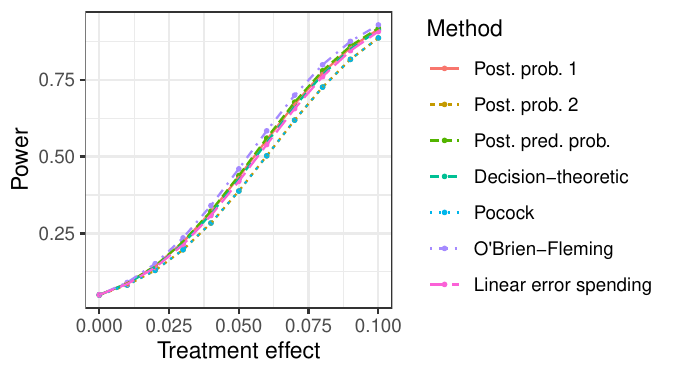}
\includegraphics[width = .83\textwidth]{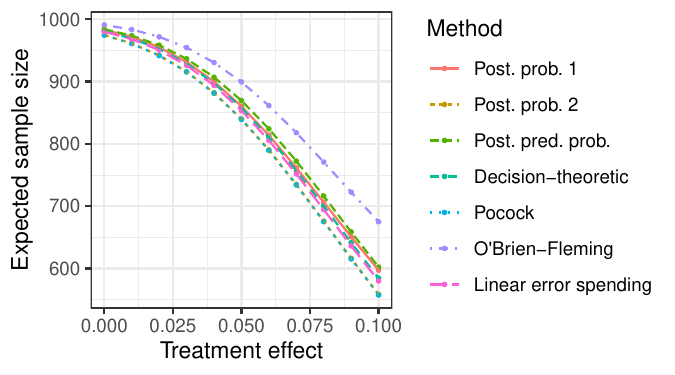}
\end{center}
\caption{Visualization of the stopping boundaries given by different sequential designs, and comparison of the frequentist properties (power and expected sample size) of the designs for hypothetical values of $\theta$, the treatment effect.}
\label{fig:boundaries}
\end{figure}

Figure \ref{fig:boundaries} shows a visualization of the stopping boundaries and a comparison of the frequentist properties of the sequential designs. Here, we consider the power and expected sample size over a range of hypothetical $\theta$ values.
There appears to be a trade-off between power and expected sample size. For example, the O'Brien-Fleming procedure has the highest power for all $\theta$ values but also requires the largest expected sample size. This is due to its large stopping boundaries at early analyses and progressively smaller stopping boundaries at later analyses.
On the contrary, the Pocock boundaries and the boundaries based on posterior probabilities (version 2) lead to the lowest expected sample size but also have the lowest power.
For more discussion on the frequentist evaluation of sequential designs, refer to \cite{jennison2000group}.

\subsection{Illustration of the Calibrated Bayesian Approach}
\label{sec:oc_Bayes_simulation}

To demonstrate the calibrated Bayesian approach, we conduct simulation studies to explore the operating characteristics of a Bayesian design under a variety of plausible scenarios.
Consider the single-arm trial example in Section \ref{sec:single_arm_example} with a maximum sample size of $n_K = 1000$ and the Bayesian design with stopping rules given by Equation \eqref{eq:bayes_PP_stopping_rule}.
Suppose the actual effect size of the trial, $\theta$, is a random draw from $\N(\mu_0, \nu_0^2)$.  As the trial progresses, patient outcomes become available sequentially and follow a normal distribution, $y_1, y_2, \ldots \sim \N(\theta, \sigma^2)$.
The trial statistician, on the other hand, uses a $\N(\mu, \nu^2)$ prior to draw inference about $\theta$, which may or may not be identical to the actual population distribution of $\theta$. For simplicity, assume the  sampling model used by the statistician, $f(\by_K \mid \theta)$, is correctly specified.
At prespecified time and frequency, the statistician conducts interim analyses of accumulating data. If the stopping rule is triggered, $H_0$ is rejected, the trial is stopped, and efficacy of the drug is declared.

We consider 72 simulation scenarios, one for each combination of $\nu_0 \in \{ 0.1, 0.5, 1\}$, $\nu \in \{ 0.1, 0.5, 1, 10 \}$, and $K \in \{ 1, 2, 5, 10, 100, 1000\}$.
For simplicity, we fix the other parameters: $\mu_0 = \mu = 0$, and $\sigma = 1$.
Here, a larger (or smaller) value of $\nu_0$ indicates that the actual effect size is more likely to be larger (or smaller). We do not consider $\nu_0 > 1$ as in practice, a standardized effect size that is much larger than what could be drawn from a $\N(0, 1^2)$ distribution is not common.
A larger (or smaller) value of $\nu$ represents that the assumed prior for $\theta$ is more diffuse (or more concentrated around zero).
When $\nu_0 = \nu$, the population distribution of $\theta$ over different trials is the same as the prior for $\theta$ used for analysis.
Lastly, $K$ is the total number of (interim and final) analyses. We assume that patients are enrolled in groups of equal size $n_K / K$.

For each scenario,  we simulate $S = 10,000$ hypothetical trials by first generating $\theta^{(1)}, \ldots, \theta^{(S)} \sim \N(\mu_0, \nu_0^2)$.
Next, for each $\theta^{(s)}$, trial outcomes are sequentially generated from $N(\theta^{(s)}, \sigma^2)$.
Interim analyses are performed after every $n_K / K$ outcomes have been observed, and the trial is stopped if the stopping rule as in Equation \eqref{eq:bayes_boundary} is satisfied with $\gamma_j \equiv \gamma = 0.95$.
We record the $\widehat{\text{FDR}}$ and $\widehat{\text{FPR}}$ as defined in Equation \eqref{eq:fdr_fpr}.
In addition, we record the percentage of 95\% credible intervals for $\theta$, calculated as in Section \ref{sec:Bayesian_analysis_completion}, that cover the true values.

\begin{table*}[ht!]
\caption{Operating characteristics of the Bayesian design with stopping rules given by Equation \eqref{eq:bayes_PP_stopping_rule}, a maximum sample size of $n_K = 1000$, $K$ planned analyses, and equal group sizes.
Values are averages over 10,000 simulated trials.
Each cell shows the corresponding metric ($\widehat{\text{FDR}}$, $\widehat{\text{FPR}}$, or Coverage) for a specific combination of $\nu_0$, $\nu$, and $K$.} 
\label{tbl:oc_Bayes_simulation}
\begin{center}
\begin{tabular}{l@{\hskip .2in}rrrr@{\hskip .3in}rrrr@{\hskip .3in}rrrr}
\toprule
$K$ & \multicolumn{4}{c}{$\widehat{\text{FDR}}$ (\%)} & \multicolumn{4}{c}{$\widehat{\text{FPR}}$ (\%)} & \multicolumn{4}{c}{Coverage (\%)} \\
\midrule
 & \multicolumn{12}{c}{$\nu_0 = 0.1$, different $\nu$ below} \\
 & \textbf{0.1} & 0.5 & 1 & 10 & \textbf{0.1} & 0.5 & 1 & 10 & \textbf{0.1} & 0.5 & 1 & 10 \\
1 & 0.8 & 0.6 & 0.8 & 0.9 & 0.5 & 0.4 & 0.5 & 0.6 & 95.0 & 95.2 & 95.3 & 94.7 \\
2 & 1.1 & 1.5 & 1.5 & 1.4 & 0.7 & 1.0 & 1.0 & 0.9 & 94.9 & 95.4 & 94.8 & 94.9 \\
5 & 1.8 & 2.8 & 3.6 & 3.1 & 1.2 & 2.0 & 2.4 & 2.1 & 94.9 & 94.7 & 94.1 & 94.5 \\
10 & 2.7 & 4.8 & 4.8 & 5.2 & 1.9 & 3.6 & 3.5 & 3.9 & 95.0 & 94.1 & 93.9 & 93.9 \\
100 & 4.2 & 11.3 & 11.7 & 12.1 & 2.9 & 9.7 & 10.3 & 10.7 & 95.1 & 93.1 & 91.8 & 91.5 \\
1000 & 5.2 & 15.1 & 19.9 & 22.5 & 3.9 & 13.5 & 19.6 & 23.5 & 95.3 & 93.7 & 91.2 & 88.1 \\
\midrule
 & \multicolumn{12}{c}{$\nu_0 = 0.5$, different $\nu$ below} \\
 & 0.1 & \textbf{0.5} & 1 & 10 & 0.1 & \textbf{0.5} & 1 & 10 & 0.1 & \textbf{0.5} & 1 & 10 \\
1 & 0.1 & 0.1 & 0.1 & 0.2 & 0.1 & 0.1 & 0.1 & 0.2 & 73.0 & 95.2 & 94.7 & 94.8 \\
2 & 0.2 & 0.3 & 0.4 & 0.1 & 0.2 & 0.3 & 0.4 & 0.1 & 67.4 & 94.9 & 94.5 & 95.3 \\
5 & 0.3 & 0.7 & 0.4 & 0.3 & 0.3 & 0.7 & 0.3 & 0.3 & 60.5 & 94.7 & 95.2 & 95.3 \\
10 & 0.6 & 0.8 & 0.8 & 0.8 & 0.5 & 0.7 & 0.7 & 0.7 & 58.3 & 95.2 & 95.0 & 95.2 \\
100 & 0.9 & 2.3 & 2.7 & 3.2 & 0.8 & 2.2 & 2.6 & 3.2 & 56.8 & 95.2 & 94.8 & 94.0 \\
1000 & 0.8 & 3.2 & 5.8 & 8.6 & 0.8 & 3.2 & 6.0 & 8.7 & 57.1 & 95.2 & 94.4 & 92.2 \\
\midrule
 & \multicolumn{12}{c}{$\nu_0 = 1$, different $\nu$ below} \\
 & 0.1 & 0.5 & \textbf{1} & 10 & 0.1 & 0.5 & \textbf{1} & 10 & 0.1 & 0.5 & \textbf{1} & 10 \\
1 & 0.0 & 0.0 & 0.0 & 0.1 & 0.0 & 0.0 & 0.0 & 0.1 & 46.8 & 94.8 & 95.1 & 94.9 \\
2 & 0.1 & 0.1 & 0.1 & 0.1 & 0.1 & 0.1 & 0.1 & 0.1 & 40.8 & 94.7 & 94.8 & 95.3 \\
5 & 0.1 & 0.2 & 0.2 & 0.2 & 0.1 & 0.2 & 0.2 & 0.2 & 36.6 & 94.4 & 95.1 & 95.0 \\
10 & 0.1 & 0.5 & 0.4 & 0.4 & 0.1 & 0.5 & 0.4 & 0.4 & 34.9 & 94.5 & 94.8 & 94.8 \\
100 & 0.3 & 1.5 & 1.3 & 1.2 & 0.3 & 1.4 & 1.3 & 1.2 & 34.2 & 90.7 & 95.1 & 94.8 \\
1000 & 0.3 & 2.2 & 3.5 & 5.1 & 0.3 & 2.2 & 3.5 & 5.3 & 33.8 & 87.6 & 94.7 & 93.4 \\
\bottomrule
\end{tabular}
\end{center}
\end{table*}

Table \ref{tbl:oc_Bayes_simulation} summarizes the simulation results.
Although the FDR and FPR increase with the number of analyses, according to Proposition \ref{prop:fdr_fpr_bound}, the FDR and FPR are upper bounded when the statistician's model is correctly specified.
These theoretical results are corroborated by the simulations: when $\nu_0 = \nu$, the $\widehat{\text{FDR}}$ is roughly bounded by $1 - \gamma = 5\%$ (due to Monte Carlo errors and a finite number of simulations, the $\widehat{\text{FDR}}$ may sometimes exceed 5\%), and the $\widehat{\text{FPR}}$ is always below $(1 - \gamma) / \gamma = 5.3\%$. 
In addition, when $\nu_0 = \nu$, the coverage of the 95\% credible intervals for $\theta$ is around 95\% regardless of $K$.

In the presence of model misspecification, however,  Bayesian statements may not attain their asserted coverage, and the discrepancy becomes larger with more frequent applications of data-dependent stopping rules.
These results are consistent with the findings in \cite{rubin1984bayesianly} and \cite{rosenbaum1984sensitivity}.
When the assumed prior is more diffuse than the actual distribution of $\theta$, the FDR and FPR are inflated, and the degree of FDR and FPR inflation becomes greater when $K$ is larger. For example, when $\nu_0 = 0.1$, $\nu = 10$, and $K = 1000$, the $\widehat{\text{FDR}}$ and $\widehat{\text{FPR}}$ are around 20\%.
For this reason, we caution against the use of diffuse  priors for decision making if data-dependent stopping rules are in frequent use and the actual effect sizes are believed to be small.
In addition, when $\nu_0 \neq \nu$, the coverage of the 95\% credible intervals for $\theta$ is below 95\% and decreases as $K$ increases.
Interestingly, an overly conservative prior (that is more concentrated around zero) results in low coverage of the credible intervals, while a diffuse prior has less impact on the coverage.

From a calibrated Bayesian point of view,  simulation studies of this type can be used to guide the choice of $\pi(\theta)$ and $\{ \gamma_1, \ldots, \gamma_K \}$.
Suppose the trial statistician decides to use a constant threshold value $\gamma_j \equiv \gamma = 0.95$ and wants to select $\nu$ such that the FDR and FPR of the design are controlled at below 5\% for plausible $\nu_0$ and $K$ scenarios (assume $\mu_0 = \mu = 0$).
To achieve this goal for all possible $\nu_0$ and $K$ considered here, $\nu$ should be set at $\leq 0.1$.
However, if one plans to conduct no more than $K = 10$ analyses, then setting $\nu \leq 1$ is sufficient.

We do not present additional numerical studies for the subjective Bayesian approach, in which case the prior and threshold values may be chosen based on a subjective belief rather than simulations.

\section{Discussion}
\label{sec:discussion}

We have summarized three perspectives on Bayesian sequential designs, namely the frequentist-oriented perspective, the subjective Bayesian perspective, and the calibrated Bayesian perspective, and have discussed their implications.
We have reviewed Bayesian sequential designs based on posterior probabilities, posterior predictive probabilities, and decision-theoretic frameworks.
We have also commented on the role of the LP in sequential trial designs.
While the LP implies that unrealized events are irrelevant to the statistical evidence about the treatment effect,  it gives little guidance in assessing a decision procedure thus does not preclude the use of additional information in decision-making.

So far, we have only considered early stopping for efficacy.  In practice, it may be desirable to allow for early stopping when interim results suggest the investigational drug is unlikely to have a clinically meaningful treatment effect \citep{snapinn2006assessment}. This is known as early stopping for futility.
A sequential trial design can include a provision for either early efficacy stopping, early futility stopping, or both.
Consider the single-arm trial example.
One could stop the trial at analysis $j$ in favor of the null hypothesis if $\Pr(\theta > 0 \mid \by_j) < \tau_j$ for some threshold $\tau_j$.
Futility stopping rules do not inflate the type I error rate; actually, they decrease the type I error rate.
However, futility stopping rules also decrease the power and increase the false negative rate (FNR) and false omission rate (FOR) of a design.
The futility boundaries could be specified to either satisfy certain power and type I error rate requirements (similar to \cite{pampallona1994group}), reflect subjective beliefs, or achieve desirable FNR, FOR,  FDR, and FPR under plausible scenarios.

Two-sided tests and point null hypotheses are very common in clinical trials. For example, for the single-arm trial in Section \ref{sec:single_arm_example}, one may test
\begin{align}
H_0: \theta = 0 \quad \text{vs} \quad H_1: \theta \neq 0.
\label{eq:test_point_null}
\end{align}
There have been several criticisms of testing a point null hypothesis \citep{berger1987testing}, such as the plausibility of $\theta$ being equal to $0$ exactly.
As a result, we have focused on a one-sided test with a composite null hypothesis (Equation \ref{eq:test}).
Most of our discussions are still applicable to tests like Equation \eqref{eq:test_point_null}, although from a Bayesian hypothesis testing perspective, the prior for $\theta$ should include a discrete mass at the location indicated by the point hypothesis.

From a frequentist perspective, the issue of type I error rate inflation (or multiplicity) can arise from repeatedly testing a single hypothesis over time, or testing multiple hypotheses simultaneously \citep{simon1994problems}.
From a subjective Bayesian perspective, however, repeated hypothesis testing is not necessarily a problem (see Section \ref{sec:subjective_bayesian_perspective}), and multiplicity adjustments are needed only when there are multiple tests.
It is worth noting that frequentist and Bayesian philosophies on multiple testing are also quite different \citep{berry1999bayesian, sjolander2019frequentist}. 

Several R packages have been developed to facilitate the use of frequentist and Bayesian sequential designs in clinical trials. These include \texttt{gsDesign} \citep{anderson2021gsdesign} and \texttt{gsbDesign} \citep{gerber2016gsbdesign}.

\clearpage

\appendix

\section{Frequentist Sequential Designs}
\label{sec:freq}

We provide a brief review of frequentist sequential designs. Consider the single-arm trial example in Section \ref{sec:single_arm_example}. The maximum type I error rate of this sequential testing procedure is given by Equation \eqref{eq:type_I_error}.
Frequentist group sequential designs are concerned with the specification of the stopping boundaries $\{ c_1, \ldots, c_K \}$ such that Equation \eqref{eq:type_I_error} holds for prespecified $\alpha$, $K$, and $\{n_1, \ldots, n_K \}$.
The solution to Equation \eqref{eq:type_I_error} is not unique, thus restrictions on the stopping boundaries have been considered.
We give some examples next.

\subsection{The Pocock and O'Brien-Fleming Procedures}
\label{sec:pocock}

In the case of equal group sizes (that is, $n_j = j g$ for some $g$),
\cite{pocock1977group} proposed to use equal stopping boundaries by setting $c_1 = \cdots = c_K = c_{\text{P}}(K, \alpha)$, while \cite{o1979multiple}  suggested decreasing boundaries with $c_j = c_{\text{OBF}}(K, \alpha) \sqrt{K / j}$. In either case, the stopping boundaries can be solved through a numerical search. 
Note that $\bz = (z_1, \ldots, z_K)^\top$ follows a multivariate normal distribution with $\E(z_j) = \theta \sqrt{n_j} / \sigma$, $\Var(z_j) = 1$, and $\Cov(z_j, z_{j'}) = \sqrt{n_j / n_{j'}}$ for $j < j'$. Therefore,
\begin{align*}
\alpha = 1 - \Phi_K(\bc; \bm 0, \bbSigma),
\end{align*}
where $\Phi_K( \cdot; \cdot, \cdot)$ is the cumulative distribution function of a multivariate Gaussian random variable,  $\bc = (c_1, c_2, \ldots, c_K)^\top$, and $\bbSigma$ is the covariance matrix of $\bz$.

\subsection{The Error Spending Approach}
\label{sec:error_spending}

\cite{slud1982two} first considered the idea of specifying the error rate spent at each analysis, 
defined as $\kappa_j = \Pr(z_1 \leq c_1, \ldots, z_{j-1} \leq c_{j-1}, z_j > c_j \mid \theta = 0)$.
This represents the probability of rejecting $H_0$ at stage $j$ but not at any previous stages, given that $\theta = 0$.
We have $\alpha = \sum_{j=1}^K \kappa_j$. 
Once the $\kappa_j$'s are specified, one can successively calculate the stopping boundaries.
\cite{gordon1983discrete} further extended this idea and suggested to use a function to characterize the rate at which the error rate is spent.
This function, denoted by $h(u)$ ($0 \leq u \leq 1$), satisfies $h(0) = 0$ and $h(1) = \alpha$.
The $\kappa_j$'s can be chosen such that $\kappa_j = h(n_j / n_K) - h(n_{j-1} / n_K)$ (with the understanding that $n_0 = 0$).
Common choices of $h(u)$ include
\begin{align*}
h_1(u) &=  \alpha \log \left( 1+ (e-1) u \right), \\
h_2(u) &= 2 - 2 \Phi \left( q_{\alpha/2} / \sqrt{u} \right), \\
h_3(u) &= \alpha u^{b} \quad \text{for} \quad b > 0.
\end{align*}
Here, $\Phi(\cdot)$ is the cumulative distribution function of the standard normal distribution, and $q_{\alpha/2} = \Phi^{-1}(1 - \alpha/2)$ is the upper $(\alpha/2)$ quantile of the standard normal distribution, $\Phi(q_{\alpha/2}) = 1 - \alpha/2$.
It has been shown that in the case of equal group sizes, $h_1(u)$ and $h_2(u)$ produce stopping boundaries similar to those given by Pocock's and O'Brien-Fleming's procedures, respectively.
Function $h_3$ is known as the power spending function and has been studied by \cite{kim1987design}.
The error spending approach introduces greater flexibility to sequential designs, as the frequency and timing of the interim analyses do not need to be specified in advance.

\subsection{Stochastic Curtailment Based on Conditional Power}
\label{sec:stochastic_curtailment}

\cite{gordon1982stochastically} proposed the idea of \textit{stochastic curtailment} that at any point in a sequential clinical trial, if the result at the end of the trial is inevitable,  the study can be terminated early.
Consider the single-arm trial example. Suppose that at the final analysis, $H_0$ will be rejected if the final $z$-statistic $z_K > q_\eta$, where $q_\eta$ is the upper $\eta$ quantile of the standard normal distribution.
Then,  at analysis $j \in \{1, \ldots, K-1\}$,  the probability that $H_0$ will be rejected upon completion of the study, given $\theta$, is given by
\begin{align*}
\CP_j(\theta) = \Pr(z_K > q_\eta \mid \theta, \by_j),
\end{align*}
where $\by_j = (y_1, \ldots, y_{n_j})$ is the vector of accumulating data up to analysis $j$.
This is known as the \textit{conditional power}.
A simple calculation shows that 
\begin{align*}
\CP_j(\theta) = 1 - \Phi \left[ \frac{\frac{q_\eta \sigma \sqrt{n_K} - n_j \bar{y}_j}{n_K - n_j} - \theta}{\sigma \sqrt{(n_K - n_j)^{-1}}}\right].
\end{align*}
If based on current data, $H_0$ will likely be rejected at the final analysis even if the investigational drug has no treatment effect ($\theta = 0$), then the trial may be stopped early.
Mathematically, one may stop the trial early if $\CP_j(0) > \gamma$ for some threshold $\gamma$. 
This is equivalent to 
\begin{align*}
z_j > q_\eta \sqrt{n_K / n_j} + q_{1 - \gamma} \sqrt{(n_K - n_j) / n_j}.
\end{align*}
If desirable, one may use different thresholds $\gamma_j$'s at different interim analyses.
An important consideration is the type I error rate of this procedure, but \cite{gordon1982stochastically} showed that the error rate is upper bounded by $\eta / \gamma$, regardless of the number of interim analyses.
Therefore, if $\eta$ and $\gamma$ are chosen such that $\eta / \gamma \leq \alpha$, the type I error rate is maintained at or below $\alpha$, even if interim analyses are conducted at arbitrary times.
The stopping boundaries based on this argument are typically conservative.
However, if the timing of the interim analyses is specified in advance, tighter stopping boundaries can be constructed by calculating the exact type I error rate numerically.

\subsection{Analysis at the Conclusion of a Sequential Trial}

Once a sequential trial has been completed, it is often of interest to construct a point estimate and a confidence interval for the treatment effect $\theta$.  Consider again the single-arm trial example. The results of the trial can be represented by a bivariate random vector $(t, z_t)$, where $t$ denotes the time of stopping,
\begin{align*}
t = 
\begin{cases}
\min \{j: z_j > c_j\}, \, &\text{if $\exists j \in \{ 1, \ldots, K\}$ s.t. $z_j > c_j$;} \\
K, \, &\text{if $z_j  \leq c_j$ for all $j$,} \\
\end{cases}
\end{align*}
and $z_t$ is the corresponding test statistic. 
Following \cite{armitage1969repeated} or \cite{jennison2000group} (Chapter 8), the density of $(t, z_t)$ is
\begin{align*}
f(t, z_t \mid \theta) = 
\begin{cases}
\tilde{f}(t, z_t \mid \theta), &\text{if $z_t > c_t$ or $t = K$;} \\
0, & \text{if $z_t \leq c_t$ and $t \in \{ 1, \ldots, K-1 \}$,}
\end{cases}
\end{align*}
where
\begin{align*}
\tilde{f}(1, z_1 \mid \theta) = \phi(z_1 - \theta \sqrt{n_1} / \sigma),
\end{align*}
and for $t = 2, \ldots, K$,
\begin{multline*}
\tilde{f}(t, z_t \mid \theta) = \int_{-\infty}^{c_{t-1}} \tilde{f}(t-1, u \mid \theta) \cdot \frac{\sqrt{n_t}}{\sqrt{n_t - n_{t-1}}} \cdot \\
\phi \left( \frac{z_t \sqrt{n_t} - u \sqrt{n_{t-1}} - (n_t - n_{t-1}) \theta / \sigma}{\sqrt{n_t - n_{t-1}}} \right) \, \d u,
\end{multline*}
%\frac{\sqrt{n_{t-1}}}{\sqrt{n_t}} u + \frac{n_t - n_{t-1}}{\sigma \sqrt{n_t}} \theta,  \frac{n_{t} - n_{t-1}}{n_t}
with $\phi( \cdot)$ denoting the standard normal density.

The sample mean estimator, $\hat{\theta} = \bar{y}_t$, is a straightforward point estimator for $\theta$. It can be shown that $\hat{\theta}$ is also the maximum likelihood estimator (MLE). However, it is known that the MLE following a sequential trial is biased,  and one may correct it by subtracting an estimate of its bias. See, e.g., \cite{whitehead1986bias} for more details.

To construct a confidence interval for $\theta$, one needs to define an ordering of the sample space \citep{tsiatis1984exact, kim1987confidence, rosner1988exact}. For example, based on the stage-wise ordering, $(t', z_{t'}')$ is above $(t, z_t)$ if either (i) $t' = t$ and $z_{t'}' > z_t$, or (ii) $t' < t$.  In this case, $(t', z_{t'}')$ is indicative of a larger value of $\theta$ compared to $(t, z_t)$.
It can be shown  that 
\begin{align*}
\Pr[ \text{Observing an outcome above $(t, z_t)$} \mid \theta ]
\end{align*}
is a continuous and monotonically increasing function of $\theta$ for every possible trial outcome $(t, z_t)$ \citep{kim1987confidence}.
Thus, one can find unique values $\theta^L$ and $\theta^U$ which satisfy
\begin{align*}
\Pr[ \text{Observing an outcome above $(t, z_t)$} \mid \theta^\text{L} ] &= \alpha / 2, \\
\Pr[ \text{Observing an outcome above $(t, z_t)$} \mid \theta^\text{U} ] &= 1 - \alpha / 2.
\end{align*}
The two equations can be solved numerically.
Then, $(\theta^\text{L}, \theta^\text{U})$ is a $100(1 - \alpha)\%$ confidence interval for $\theta$.

\section{The Calibrated Bayesian Perspective}
\label{sec:app:calibrated_Bayes}

We present more details about the calibrated Bayesian perspective described in Section \ref{sec:calibrated_bayesian_perspective}. We consider the setup of an infinite series of single-arm trials (described in Section \ref{sec:single_arm_example}) with true but unknown treatment effects $ \theta^{(1)}, \theta^{(2)}, \ldots  \sim \pi_0(\theta)$.
For each trial, patient outcomes $\by_K \sim f_0(\by_K \mid \theta)$ and are observed sequentially.
The Bayesian design with stopping rules given by Equation \eqref{eq:bayes_PP_stopping_rule} is applied to every trial with a prior model $\pi(\theta)$, a sampling model $f(\by_K \mid \theta)$, and threshold values $\{ \gamma_1, \ldots,  \gamma_K \}$.
We are interested in the operating characteristics of the Bayesian design over this infinite series of trials, in particular its FDR and FPR.

\subsection{Background}
\label{sec:app:background_calibrated_Bayes}

We first provide more background on the calibrated Bayesian perspective. 
\cite{rubin1984bayesianly} called a statistical procedure  
(conservatively) \textit{calibrated} if the resulting probability statements (at least) have their asserted coverage in repeated practices.
Clearly, calibrated procedures are desirable, and Rubin recommended examining operating characteristics to select calibrated Bayesian procedures. Rubin's points were echoed by \cite{little2006calibrated}.

The following discussion is adopted from \cite{rubin1984bayesianly}. 
A Bayesian procedure is calibrated if the model specification is correct, that is, if $f(\by_K \mid \theta) \pi(\theta) = f_0(\by_K \mid \theta) \pi_0(\theta)$.
For example, suppose that $I(\by_K)$ is a 95\% credible interval for $\theta$ under model $f(\by_K \mid \theta) \pi(\theta)$, then
\begin{align*}
\frac{\int_{\theta \in I(\by_K)} f(\by_K \mid \theta) \pi(\theta) \d \theta}{\int_{\theta} f(\by_K \mid \theta) \pi(\theta) \d \theta} = \frac{\int_{\theta \in I(\by_K)} f_0(\by_K \mid \theta) \pi_0(\theta) \d \theta}{\int_{\theta} f_0(\by_K \mid \theta) \pi_0(\theta) \d \theta} = 0.95.
\end{align*}
The interpretation is that, among the possible $\theta$ values from $\pi_0(\theta)$ that might have generated the observed $\by_K$ from $f_0(\by_K \mid \theta)$, 95\% of them belong to $I(\by_K)$. 
Therefore, when the procedure of calculating $I(\by_K)$ from $f(\by_K \mid \theta) \pi(\theta)$ is repeatedly applied to data drawn from $f_0(\by_K \mid \theta) \pi_0(\theta)$,  95\% of the calculated credible intervals will cover the true parameter values.
We see that posterior probabilities correspond to frequencies of actual events.
Similarly, when we claim $\Pr(\theta > 0 \mid \by_K) > 0.95$, it means that among the possible $\theta$ values that might have generated $\by_K$, more than 95\% are positive.

\cite{rubin1984bayesianly} and \cite{rosenbaum1984sensitivity} also demonstrated that when the model specification is correct, the coverage and interpretation of Bayesian statements are still valid under data-dependent stopping rules.
For example, if we conclude $\Pr(\theta > 0 \mid \by_j) > 0.95$ at any interim analysis $j$, it means that more than 95\% of the possible $\theta$ values that might have generated $\by_j$ are positive, even if the trial is optionally stopped at analysis $j$ based on the observed data.

Of course, in the presence of model misspecification, the coverage of Bayesian statements is not warranted.
In particular, \cite{rubin1984bayesianly} and \cite{rosenbaum1984sensitivity} noted that data-dependent stopping rules increase the sensitivity of Bayesian inference to model specification.
Therefore, especially for sequential trial designs, one might want to examine their operating characteristics for a range of plausible $f_0(\by \mid \theta) \pi_0(\theta)$ (which may deviate from $f(\by \mid \theta) \pi(\theta)$) to select appropriate design parameters.

\subsection{The False Discovery Rate}
\label{sec:app:fdr}

We show that the FDR is upper bounded if $f(\by_K \mid \theta) \pi(\theta) = f_0(\by_K \mid \theta) \pi_0(\theta)$.
Note that if $\by_K \in \Gamma$, then $\Pr(\theta > 0 \mid \by_K) > \gamma_{\min}$. This is because for every $j \in \{ 1, \ldots, K \}$,
\begin{align*}
\Pr(\theta > 0 \mid \by_j) = \int_{\by_{j, K}} \Pr(\theta > 0 \mid \by_j, \by_{j, K}) f(\by_{j, K} \mid \by_j) \d \by_{j, K},
\end{align*}
where $\by_{j, K} = (y_{n_j+1}, \ldots, y_{n_K})$.
If $\Pr(\theta > 0 \mid \by_K) = \Pr(\theta > 0 \mid \by_j, \by_{j, K}) \leq \gamma_{\min}$, then $\Pr(\theta > 0 \mid \by_j) \leq \gamma_{\min}$ for every $j$, which contradicts with $\by_K \in \Gamma$.
Therefore, 
\begin{align*}
\text{FDR} &= \frac{\int_{\by_K \in \Gamma} \int_{\theta \leq 0}  f_0(\by_K \mid \theta) \pi_0(\theta) \d \theta \d \by_K}{\int_{\by_K \in \Gamma} f_0(\by_K) \d \by_K} \\
&= \frac{\int_{\by_K \in \Gamma} \int_{\theta \leq 0}  f(\by_K \mid \theta) \pi(\theta) \d \theta \d \by_K}{\int_{\by_K \in \Gamma} f(\by_K) \d \by_K} \\
&= \frac{\int_{\by_K \in \Gamma} \Pr(\theta \leq 0 \mid \by_K) \cdot f(\by_K) \d \by_K}{\int_{\by_K \in \Gamma} f(\by_K) \d \by_K} \\
&\leq 1 - \gamma_{\min}.
\end{align*}

\subsection{The False Positive Rate}
\label{sec:app:fpr}

To derive the upper bound of the FPR when $f(\by_K \mid \theta) \pi(\theta) = f_0(\by_K \mid \theta) \pi_0(\theta)$, we first introduce an inequality under the Bayesian hypothesis testing framework (Section \ref{sec:Bayesian_hypothesis_testing}).
Assume 
\begin{align*}
\theta \mid H_0 \sim \pi^{(0)}(\theta), \qquad 
\theta \mid H_1 \sim \pi^{(1)}(\theta),
\end{align*}
and write $f(\by_j \mid H_m) = \int_{\theta} f(\by_j \mid \theta) \pi^{(m)}(\theta) \d \theta$ for $j = 1, \ldots, K$ and $m = 0, 1$. Then, the following inequality holds for any $0 < \epsilon < 1$ \citep{hendriksen2021optional}:
\begin{align*}
\Pr \left[  \exists j \in \{ 1, \ldots, K \}: \frac{f(\by_j \mid H_1)}{f(\by_j \mid H_0)} > \frac{1}{\epsilon} 
\,\middle\vert\,   
H_0 \right] \leq \epsilon,
\end{align*}
where $\Pr(\cdot \mid H_0) = \int_{\theta} \Pr(\cdot \mid \theta) \pi^{(0)}(\theta) \d \theta$.
This is referred to as a \textit{universal bound} on the probability of observing misleading evidence \citep{royall2000probability, sanborn2014frequentist}.

In our application, instead of specifying the priors for $\theta$ separately under $H_0$ and $H_1$, a single prior for $\theta$ is specified over the entire parameter space,  $\theta \sim \pi(\theta)$. 
Still, the universal bound is applicable,  because $\theta \sim \pi(\theta)$ is equivalent to
\begin{align*}
&\Pr(H_0) =  \int_{\theta \leq 0} \pi(\theta) \d \theta,  \quad \Pr(H_1) =  \int_{\theta > 0} \pi(\theta) \d \theta, \\ 
&\theta \mid H_0 \sim \pi(\theta \mid \theta \leq 0) = \frac{\pi(\theta) \cdot \bm 1 (\theta \leq 0)}{\int_{\theta \leq 0} \pi(\theta) \d \theta},  \\
&\theta \mid H_1 \sim \pi(\theta \mid \theta > 0) =  \frac{\pi(\theta) \cdot \bm 1 (\theta > 0)}{\int_{\theta > 0} \pi(\theta) \d \theta}.
\end{align*}
Also, $\Pr(\theta > 0 \mid \by_j) = \Pr(H_1 \mid \by_j) > \gamma_j$ is equivalent to 
\begin{align*}
\frac{f(\by_j \mid H_1)}{f(\by_j \mid H_0)} > \frac{\gamma_j \cdot \int_{\theta \leq 0} \pi(\theta) \d \theta}{(1 - \gamma_j) \cdot \int_{\theta > 0} \pi(\theta) \d \theta}.
\end{align*}
Applying the universal bound and notice that $f(\by_K \mid \theta) \pi(\theta) = f_0(\by_K \mid \theta) \pi_0(\theta)$, we have
\begin{align*}
\text{FPR} &= \frac{\int_{\by_K \in \Gamma}  \int_{\theta \leq 0}  f_0(\by_K \mid \theta) \pi_0(\theta) \d \theta \d \by_K }{\int_{\theta \leq 0} \pi_0(\theta) \d \theta} \\
&= \int_{\theta} f (  \by_K \in \Gamma \mid \theta ) \pi (\theta \mid \theta \leq 0) \d \theta \\
&= \Pr \Bigg[  \exists j \in \{ 1, \ldots, K \}: \frac{f(\by_j \mid H_1)}{f(\by_j \mid H_0)} > \frac{\gamma_j \cdot \int_{\theta \leq 0} \pi(\theta) \d \theta}{(1 - \gamma_j) \cdot \int_{\theta > 0} \pi(\theta) \d \theta}
\,\Bigg\vert\,   
H_0 \Bigg] \\
&\leq \Pr \Bigg[  \exists j \in \{ 1, \ldots, K \}: \frac{f(\by_j \mid H_1)}{f(\by_j \mid H_0)} > \frac{\gamma_{\min} \cdot \int_{\theta \leq 0} \pi(\theta) \d \theta}{(1 - \gamma_{\min}) \cdot \int_{\theta > 0} \pi(\theta) \d \theta}
\,\Bigg\vert\,   
H_0 \Bigg] \\
&\leq \frac{(1 - \gamma_{\min}) \cdot \int_{\theta > 0} \pi(\theta) \d \theta}{\gamma_{\min} \cdot \int_{\theta \leq 0} \pi(\theta) \d \theta}.
\end{align*}

\clearpage

\bibliographystyle{apalike}
\bibliography{Sequential_designs_ref}

\end{document}